\documentclass{aa}

\usepackage{txfonts}
\usepackage{graphics,epsfig}
\usepackage{graphicx}
\usepackage{amsmath}
\usepackage{amssymb}
\usepackage{rotating}
\usepackage{ulem}
\usepackage{float}
\usepackage{color}
\usepackage{appendix}
\usepackage{caption}
\usepackage{natbib}
\usepackage[hyperindex,breaklinks=true, colorlinks, citecolor=blue]{hyperref}  
\bibpunct{(}{)}{;}{a}{}{,}

\newcommand{\kms}{\rm km~s$^{-1}$}
\newcommand{\acc}{\ensuremath{\rm atoms~cm^{-2}}}
\newcommand{\arcs}{$^{\prime\prime}$}
\newcommand{\arcm}{$^{\prime}$}

\title{A first quantification of the effects of absorption for H{\sc i} Intensity Mapping experiments}

\titlerunning{H{\sc i} absorption in Intensity Mapping}

\author{Sambit Roychowdhury\inst{1,2,3}, Clive Dickinson\inst{1}, Ian W. A. Browne\inst{1}}

\institute{$^1$ Jodrell Bank Centre for Astrophysics, Alan Turing building, Department of Physics \& Astronomy, School of Natural Sciences, The University of Manchester, Oxford Road, Manchester M13 9PL, UK\\
$^2$ Institut d'Astrophysique Spatiale, CNRS, Université Paris-Sud, Université Paris-Saclay, Bât. 121, 91405 Orsay Cedex, France\\
$^3$ Centre for Astrophysics and Supercomputing, Swinburne University of Technology, Hawthorn, VIC 3122, Australia\\
\email{sroychowdhury@swin.edu.au}
}

\authorrunning{Roychowdhury et al.}

\date{Received; accepted}

\begin{document}

\abstract
{H{\sc i} Intensity Mapping (IM) will be used to do precision cosmology, using many existing and upcoming radio observatories. It will measure the integrated H{\sc i} 21\,cm emission signal from `voxels' of the sky at different redshifts. The signal will be  contaminated due to absorption, the largest component of which will be the flux absorbed by the H{\sc i} emitting sources themselves from the potentially bright flux incident on them from background radio continuum sources.}
{We, for the first time, provide a quantitative estimate of the magnitude of the absorbed flux compared to the emitted H{\sc i} flux. The ratio of the two fluxes was calculated for various voxels placed at redshifts between 0.1 and 2.5.}
{We used a cosmological sky simulation of the atomic H{\sc i} emission line, and summed over the emitted and absorbed fluxes for all sources within voxels at different redshift. In order to determine the absorbed flux, for each H{\sc i} source the flux incident from background radio continuum sources was estimated by determining the numbers, sizes, and redshift distribution of radio continuum sources that lie behind it, based on existing observations and simulations. The amount of this incident flux that is absorbed by each H{\sc i} source was calculated using a relation between integrated optical depth with H{\sc i} column density determined using observations of damped Lyman-$\alpha$ systems (DLAs) and sub-DLAs.}
{We find that for the same co-moving volume of sky, the H{\sc i} emission decreases quickly with increasing redshift, while the absorption varies much less with redshift and follows the redshift distribution of faint sources that dominate the number counts of radio continuum sources. This results in the fraction of absorption compared to emission to be negligible in the nearby Universe (up to a redshift of $\sim$0.5), increases to about 10\% at a redshift of one, and continues to increase to about 30\% up to a redshift of 2.5. These numbers can vary significantly due to the uncertainty on the exact form of the following relations: firstly, the number counts of radio continuum sources at sub-mJy flux densities; secondly, the relation between integrated optical depth and H{\sc i} column density of H{\sc i} sources; and thirdly, the redshift distribution of radio continuum sources up to the highest redshifts.}
{Absorption of the flux incident from background radio continuum sources might become an important contaminant to H{\sc i} IM signals beyond redshifts of 0.5. 
 The impact of absorption needs to be quantified more accurately using inputs from upcoming deep surveys of radio continuum sources, H{\sc i} absorption, and H{\sc i} emission with the Square Kilometre Array and its precursors.}


\maketitle

\section{Introduction}
\label{sec:int}

H{\sc i} Intensity Mapping (IM) refers to a novel method of doing precision cosmology using the integrated H{\sc i} 21\,cm emission signal from volumes or `voxels' of the sky at different redshifts, starting immediately after reionization right up to the present day \citep{2004MNRAS.355.1339B,2006astro.ph..6104P,2008PhRvL.100p1301L,2014ApJ...781...57S,2014IAUS..306..165S,2015ApJ...803...21B,2017MNRAS.470.3220W,2017arXiv170909066K,2018ApJ...866..135V}.
The cosmic H{\sc i} signal is weak and needs to be integrated over large volumes of sky for it to be detectable as a fluctuation, over and above, the statistical and instrumental noise.
There are also much brighter `foregrounds' that need to be removed for the H{\sc i} signal to become detectable -- radio synchrotron and free-free emissions from the Galaxy itself, radio continuum emission from extragalactic sources, instrumental effects like $1/f$ noise, etc.
In spite of these challenges H{\sc i} IM promises to be one of the most important tools to study the evolution of large scale structure and consequently cosmology.
There are multiple ongoing efforts to use existing \citep{2009MNRAS.394L...6P,2010Natur.466..463C,2010PhRvD..81j3527M,2013ApJ...763L..20M,2014IAUS..306..165S,2018MNRAS.476.3382A} and upcoming radio observatories \citep{2008PhRvL.100i1303C,2012IJMPS..12..256C,2013MNRAS.434.1239B,2014SPIE.9145E..22B,2016ASPC..502...41B,2016SPIE.9906E..5XN,2016arXiv161006826B,2017A&A...597A.136S} to carry out H{\sc i} IM.
Cosmology using H{\sc i} IM is one of the main science objectives of the Square Kilometre Array \citep[SKA,][]{2015aska.confE..19S,2017MNRAS.466.2736V,2018arXiv181102743S}, and there is a proposal to perform an H{\sc i} IM survey with the SKA precursor, MeerKAT \citep{2017arXiv170906099S}.

The success of H{\sc i} IM will depend on precisely measuring integrated signals from the sky.
We therefore need to understand and estimate all possible contributions that can contribute to the integrated signal, their comparative strengths, and work out methods to extract the H{\sc i} emission signal from the observed integrated signal.
There exist estimates of the measured flux that will be detected by H{\sc i} IM experiments for observations centred at various redshifts with varying configurations \citep[e.g.][]{2015MNRAS.454.3240B,2018MNRAS.478.2416H}.
These estimates take into account various `foregrounds' as well as instrumental effects, qnd show that sky and instrumental foreground signals are expected to dominate the integrated signal.
The foreground signals correlate in contrast to the H{\sc i} signal \citep{2012A&A...540A.129A,2012MNRAS.419.3491L}, and various methods have been developed to exploit this fact and `subtract' the foreground signals\citep[e.g.][]{2011PhRvD..83j3006L,2013ApJ...763L..20M,2013MNRAS.434L..46S,2014ApJ...781...57S,2014MNRAS.441.3271W,2015ApJ...814..145A,2015ApJ...815...51S,2016MNRAS.456.2749O}, and thus extract the H{\sc i} emission signal.

\begin{figure}
\begin{center}
\includegraphics[width=3.6truein]{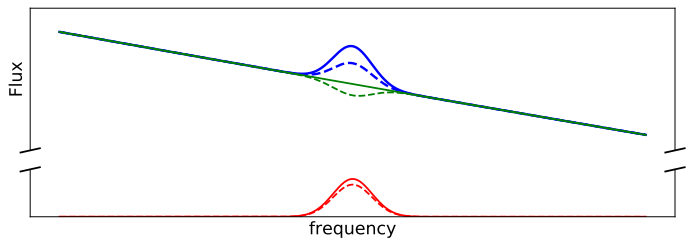}
\end{center}
\caption{Schematic representation of how absorption will affect the signal that an H{\sc i} IM experiment aims to detect. The total H{\sc i} emission from sources within a given volume of the sky is represented by the {\it bold red line}, where the frequency spread corresponds to the redshift spread of the H{\sc i} emitting sources within the volume. The {\it dashed red line} represents the H{\sc i} emission after correcting for any potential absorption within the emitting sources themselves and within sources in the foreground. The {\it bold green line} represents the broadband spectrum of the radio continuum emission from sources in the foreground as well as the background of the H{\sc i} emitting sources. We note that the flux level of this emission, which is one of the foregrounds for H{\sc i} IM, is likely to be much higher than the H{\sc i} emission one aims to detect. The {\it dashed green line} represents the radio continuum emission corrected for the absorption of a fraction of the emission incident from background radio continuum sources by the H{\sc i} emitting sources themselves. The {\it bold blue line} represents the total of the H{\sc i} emission and radio continuum emission (sum of the bold red and green lines), where the H{\sc i} emission can be thought to be a positive fluctuation on top of a spectrally smooth broadband radio continuum `foreground'. The {\it dashed blue line} represents how the actual combined H{\sc i} and radio continuum emission signal will look like, after correcting for the negative fluctuation due to the various absorptions (sum of the dashed red and green lines).}
\label{fig:sch}
\end{figure}

IM experiments aim to detect the combined H{\sc i} 21\,cm emission from individual H{\sc i} rich sources within a volume of sky.
This emission from the (unresolved) H{\sc i} sources can be visualized as a fluctuation on top of a base flux level having a smooth {broadband} spectrum due to the various `foregrounds', {as shown in the schematic Fig.~\ref{fig:sch}.
The astronomical `foregrounds' in H{\sc i} IM will consist of radio continuum emission from our Galaxy, and radio continuum emission from extragalactic sources in front of the H{\sc i} emitting sources we are interested in (i.e. $z<z_{IM}$ where $z_{IM}$ is the redshift of an H{\sc i} emitting source), as well as extragalactic sources in the background (i.e. $z>z_{IM}$).
A part of the H{\sc i} flux being emitted will be absorbed within the H{\sc i} emitting sources themselves, and also by sources in the foreground including the Galaxy.
More importantly, the H{\sc i} in the sources of interest will absorb a part of the emission incident on them from background radio continuum sources.
Therefore, along with the positive fluctuation due to H{\sc i} emission there will be some negative fluctuation due to the combined effect of these various absorptions.
This negative fluctuation will also enter the combined signal being picked up by the IM experiment, similar to what is shown in Fig.~\ref{fig:sch}.}
To date there has been no attempt to quantify the amount of absorption, the assumption being that the absorption would be low compared to the emission.

In order to understand the problem at hand, the basic equation to consider is the brightness temperature $T_B$ of an isothermal cloud of atomic hydrogen in front of a background source of brightness temperature $T_C$,
\begin{equation}
\label{eqn:bas}
T_B(v)~=~T_S~(1-e^{-\tau(v)})~+~T_C(v)~e^{-\tau(v)},
\end{equation}
where $\tau(v)$ is the optical depth and $T_S$ is the spin temperature of the hydrogen in the cloud.
In practice, the brightest radio continuum sources in the field would be flagged based on radio continuum source catalogues.
For the fainter uncatalogued radio continuum sources, their base flux level would be determined and removed during `foreground subtraction' using the smooth spectral slope of radio continuum sources.
Thus at the redshifted frequency of H{\sc i} 21\,cm emission from a given H{\sc i} source being observed by the IM experiment, the base flux level calculated for any background source is what it would have been without any absorption by the H{\sc i} source itself.
Therefore the flux absorbed by the H{\sc i} source itself becomes part of the H{\sc i} `signal'. 
Therefore effectively the brightness temperature that will be measured is,
\begin{equation}
\label{eqn:bas1}
T_B(v)~=~T_S~(1-e^{-\tau(v)})~-~T_C(v)~(1-e^{-\tau(v)}).
\end{equation}
Analytical estimates of the brightness temperature $T_B$ of the H{\sc i} emission that would be observable by IM experiments \citep[e.g. as in][]{2013MNRAS.434.1239B} ignore the term related to $T_C$.
The spin temperature for hyperfine 21\,cm emission line of H{\sc i} in astrophysical situations is set by collisional excitation of the hydrogen atoms.
Thus working under the assumption that radiative excitation of hydrogen atoms is negligible, the physics of radiative transfer for stimulated emission can be used to find an expression for $T_S(N_{HI})$, where $N_{HI}$ is the column density of atomic hydrogen in the cloud \citep{1958PIRE...46..240F,1978ApJS...36...77D}.
Analytical estimates of the brightness temperature also assume that the H{\sc i} emitting gas is optically thin.
Eqn. \ref{eqn:bas1} therefore becomes,
\begin{equation}
\label{eqn:bas2}
T_B(v)~=~T_S(N_{HI})~\tau(v).
\end{equation}
Estimates based on the latest hydrodynamical simulations \citep[e.g.][]{2018ApJ...866..135V} start with modelling the amount and distribution of H{\sc i} in detail using state-of-the-art prescriptions.
But when converting the column densities of H{\sc i} thus derived into a brightness temperature, they also work under the assumption that the H{\sc i} emitting gas is optically thin, and do not take into account the terms related to $T_C$ in eqn.~\ref{eqn:bas1}.
{The validity of the assumption about the H{\sc i} being optically thin and the potential correction for H{\sc i} self-absorption are discussed in Section~\ref{ssec:self}.}

Analytical estimates show that the rms fluctuation corresponding to the emission term $T_S~(1-e^{-\tau(v)})~/~T_S~\tau(v)$, within the power spectrum that will be measured by an H{\sc i} IM experiment will be $\sim$0.1 mK \citep{2013MNRAS.434.1239B}.
Correspondingly, the rms fluctuation due to radio continuum sources (with sources $>$10 mJy flagged), or $T_C$, will be $\sim$10 mK.
Therefore it is important to confirm whether in  eqn.~\ref{eqn:bas1} the factor due to the optical depth offered by the H{\sc i} sources ($1-e^{-\tau(v)}$) will be small enough so that the second term on the right hand side is negligible compared to the first.
We will focus exclusively on this type of absorption in this paper.
Other absorptions are possible but expected to have negligible effect.
See Appendix~\ref{app:abs} for a more complete discussion.

It has been proposed that the large number of H{\sc i} absorption systems to be detected by upcoming H{\sc i} IM surveys can potentially be used to statistically measure the Sandage-Loeb effect by observing the systems over a long period of time, and thus directly measure the accelaration of the universe \citep{2014PhRvL.113d1303Y,2019arXiv190501184J}.
In this paper though we focus on how significant the flux absorbed from background radio continuum sources by the H{\sc i} emitting sources is, when compared to the emitted H{\sc i} flux that H{\sc i} IM experiments ultimately aim to detect.
The paper is organized as follows.
In Section~\ref{sec:cal} we detail the method and the underlying assumptions that we use to calculate the amount of such absorption. 
In Section~\ref{sec:res} we present our estimates for the absorption flux compared to the emission flux for all sources integrated over a voxel in the sky.
We discuss the significance of our results, and how the various assumptions might affect our results.
Finally in Section~\ref{sec:con} we summarize our results and what needs to be done in the future to fully understand the effects of absorption on H{\sc i} IM experiments.

\section{Method}
\label{sec:cal}

\begin{figure*}
\begin{center}
\begin{tabular}{cc}
{\mbox{\includegraphics[width=3.5truein]{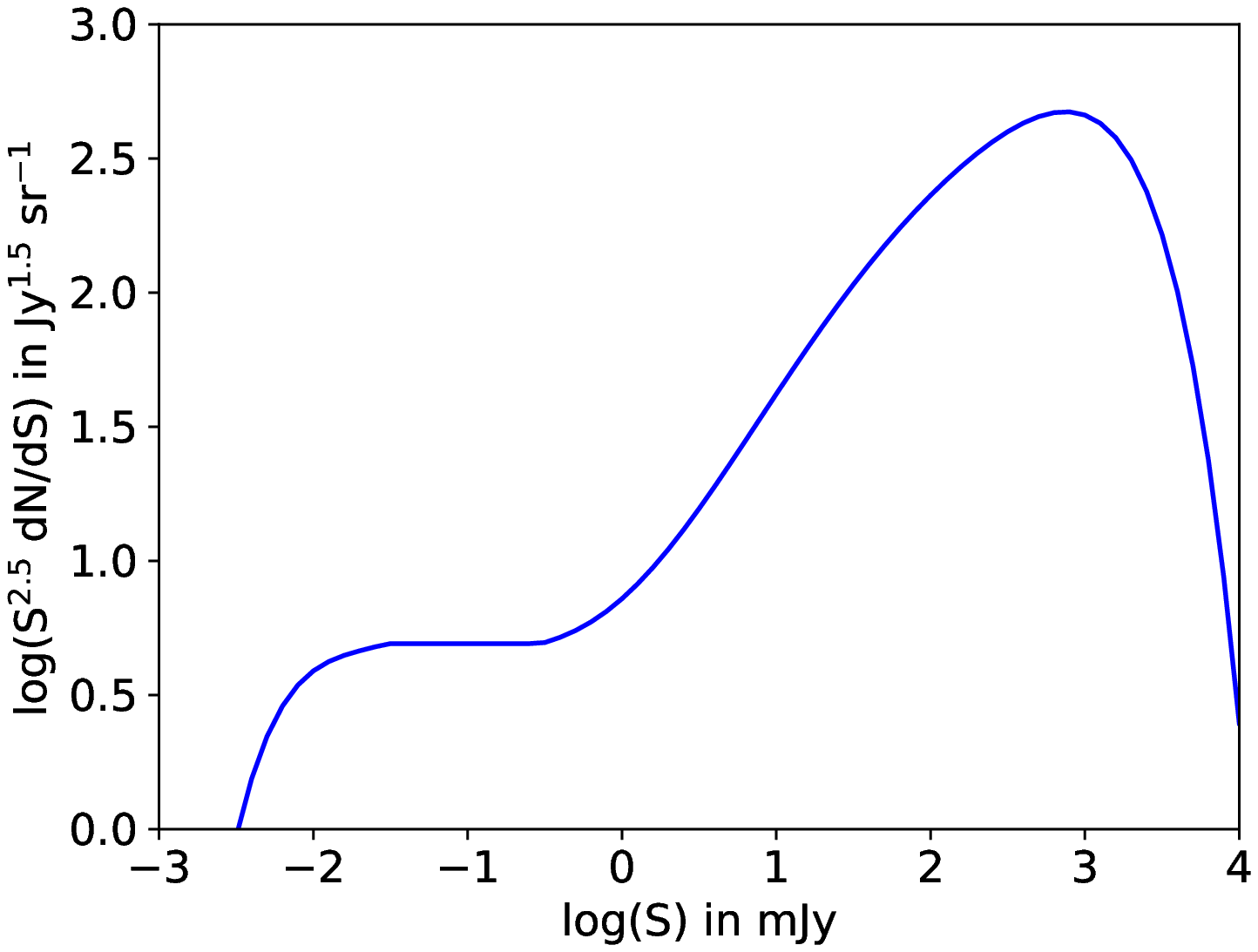}}}&
{\mbox{\includegraphics[width=3.5truein]{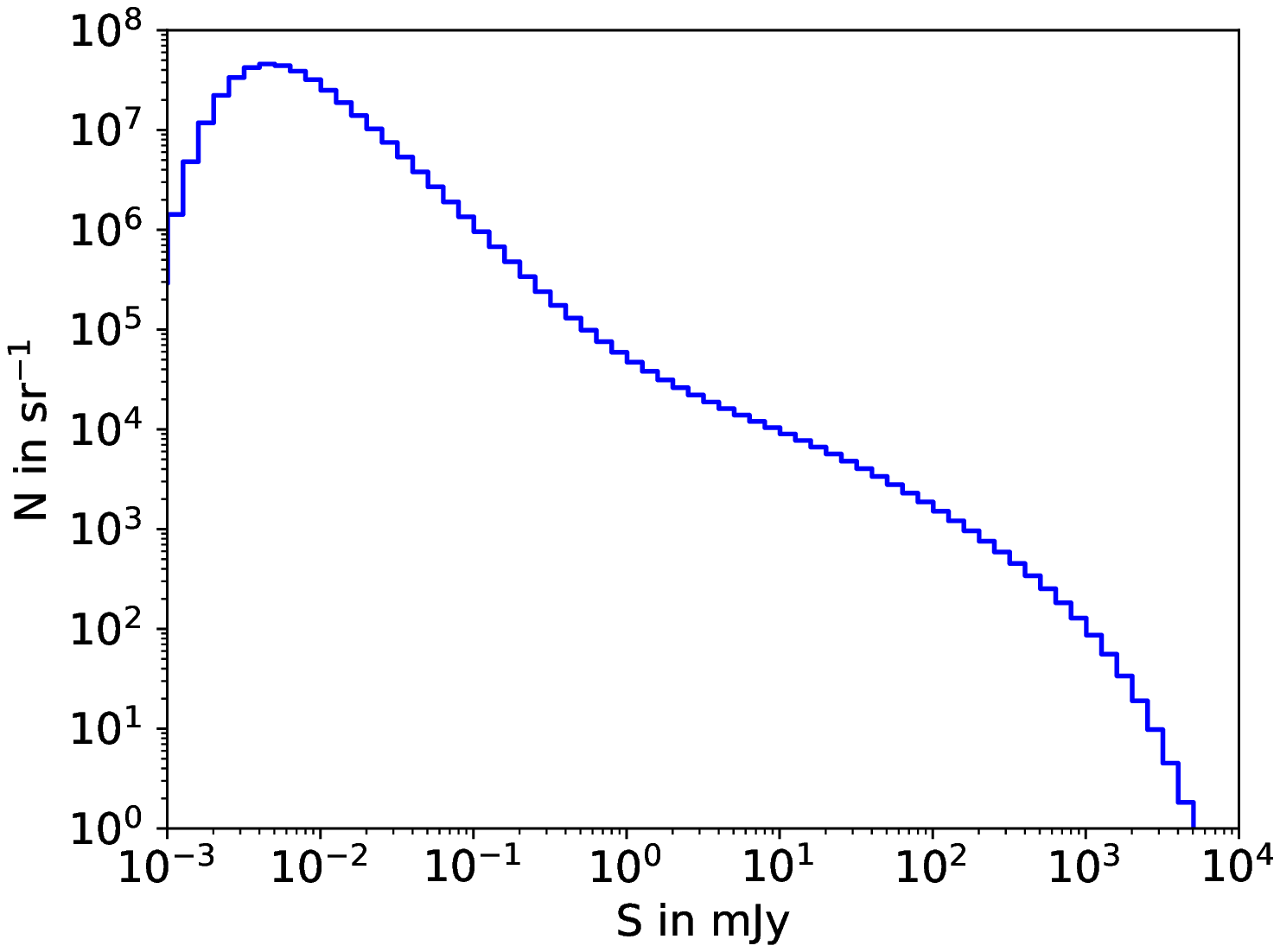}}}\\
\end{tabular}
\end{center}
\caption{The number count of sources at 1.4 GHz \citep{2014MNRAS.441.2555H}: Euclidean normalized differential ({\it left panel}) and total ({\it right panel}) as functions of flux density.}
\label{fig:dnds}
\end{figure*}

\begin{figure}
\begin{center}
\begin{tabular}{c}
{\mbox{\includegraphics[width=3.5truein]{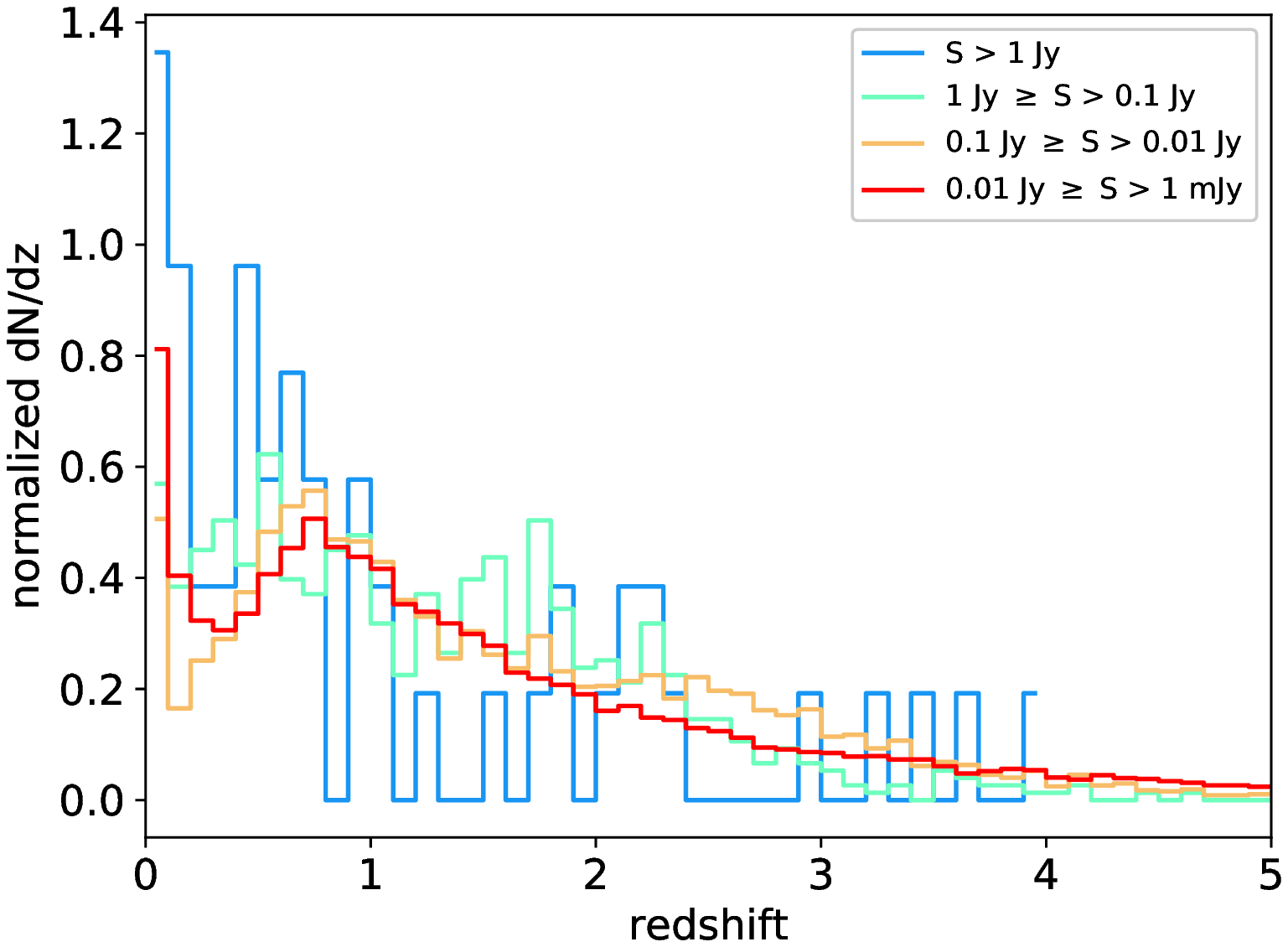}}}\\
{\mbox{\includegraphics[width=3.5truein]{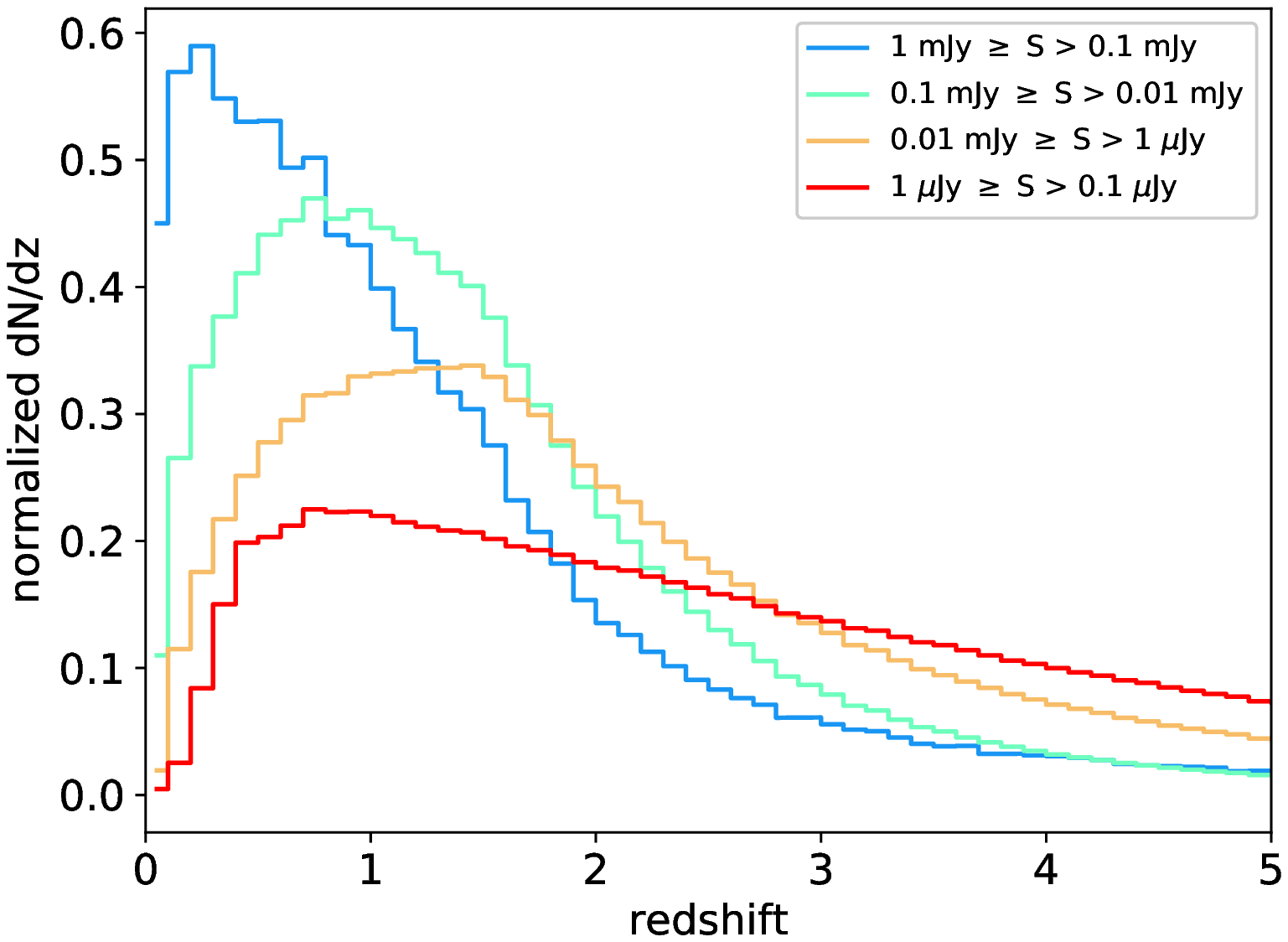}}}
\end{tabular}
\end{center}
\caption{The simulated redshift distribution of radio continuum sources in various flux density bins, from \citep{2008MNRAS.388.1335W}. The upper panel are for fluxes above 1 mJy and has a y-axis range which is more than twice as large as the lower panel.}
\label{fig:contz}
\end{figure}

From this point in the paper we focus on what is likely to be the most significant type of absorption -- that of incident radiation from background continuum sources  intercepted in the H{\sc i} clouds (sources) responsible for the H{\sc i} emission we are interested in detecting.
For a given IM experiment observing a given voxel in the sky, our aim was to calculate the ratio of the total flux absorbed from the incident background continuum, to the H{\sc i} 21\,cm flux emitted by the H{\sc i} sources within the voxel.
This ratio will provide a quantitative measure of the significance of absorption in H{\sc i} IM experiments, as in its denominator is the cosmological signal one is actually interested in measuring.
We used the cosmological sky simulation of the atomic H{\sc i} emission line by \citet{2009ApJ...698.1467O,2009ApJ...703.1890O}, hereafter the O9 simulation/catalogue, for our estimates.
{The simulation is built by post-processing a semi-analytic virtual galaxy catalogue \citep{2007MNRAS.375....2D} built on the Millenium Simulation of cosmic structure.
In the O9 simulations the cold gas is split into H{\sc i} and H$_2$, spatial distribution and velocity profiles are assigned separately to the atomic and molecular gas components, and the resulting galaxies compared against the observed mass functions and various scaling relations.
This is important as we want to focus exclusively on the H{\sc i} properties of the sources.
These simulations are ideal for large-scale cosmological predictions where individual galaxies are represented by a handful of global properties.
As we detail below, a simple representation of the H{\sc i} disks in galaxies is what we used for our present estimates, as any further details become redundant given the large uncertainties that exist for other observables and properties that need to be factored into our estimates.}
Various parameters for H{\sc i} sources including positions, gas contents, morphologies, velocity information, etc. are available up to a redshift of $z_{max}~=~10$.
For reasons detailed in Section~\ref{ssec:od}, we restricted our calculation to below redshift $z=2.5$.
This is also a practical limit given that $z\sim2.5$ is the upper limit of even the most ambitious targeted H{\sc i} IM experiments, and approaching the upper limit ($z\sim3$) of planned H{\sc i} IM surveys with the SKA \citep{2015ApJ...803...21B}.

{For each each H{\sc i} source at a given redshift, we estimated the number of radio continuum sources in different flux density ranges that will lie behind it.
We also estimated the covering factor (fraction of the background source that is covered by the foreground source) for any radio continuum source given its size on the sky based on its flux density.}
The fraction of the incident flux from all these continuum sources that is absorbed by the H{\sc i} source is then calculated based on the optical depth of the H{\sc i} in the source.
Finally, the total fluxes emitted and absorbed by all H{\sc i} sources within a given voxel of sky is summed up.

The various intermediate steps required to ultimately calculate the total emitted and absorbed fluxes from the H{\sc i} sources within a voxel of the O9 simulated sky are listed in the sub-sections below.
In many of these steps we had to make some assumptions and simplifications in order to calculate the relevant quantities.
We discuss the validity and shortcomings of each assumption either in this section or later in Section~\ref{sec:res}.

\subsection{H{\sc i} source parameters}
\label{ssec:his}

We started with a couple of assumptions about the H{\sc i} sources themselves.
Firstly, the H{\sc i} in the source is distributed uniformly over an elliptical disk.
This is a valid approximation even when nearby H{\sc i} rich galaxies are observed, but at low resolution.
For H{\sc i} IM experiments {given the scales of interest}, the angular resolution would not be good enough even to resolve the disk of any H{\sc i} source being observed, keeping in mind we are interested in sources at cosmological distances and not in the local universe.\\
Secondly, we assumed that the H{\sc i} is distributed uniformly over some velocity width.
This is a workable assumption given the double-horned profile of the H{\sc i} emission used in the O9 simulations.
We should be aware though that unlike spatial resolution, H{\sc i} IM experiments would be observing with high spectral resolution.
This might change the results of our calculation slightly for say a narrow channel width centred on the peak of the H{\sc i} emission.
In the end though, in order to have enough signal-to-noise for constraining cosmology the bandwidths would be much larger than the observing channel width \citep[e.g. see][]{2013MNRAS.434.1239B} -- thus this assumption too would be valid.

\subsection{Number of background sources per unit flux density}
\label{ssec:dnds}

We needed to know the number and distribution in terms of spectral flux density, of radio continuum sources that would be covered by the H{\sc i} source of interest.
In order to do so, we started with the observed number count of radio sources at 1.4 GHz, the assumption being that the distribution is the same at the frequency corresponding to that of H{\sc i} 21\,cm emission at the redshift of the H{\sc i} source.
\citet{2014MNRAS.441.2555H} provide an empirical fit to the number count of radio sources as observed in the second data release of the Australia Telescope Large Area Survey (ATLAS) at 1.4 GHz done with the Australia telescope Compact Array (ATCA), using the parametric form for the fit to the Phoenix Deep Survey and Faint Images of the Radio Sky at Twenty cm (FIRST) survey data by \citet{2003AJ....125..465H} while accounting for the observed excess in the number of radio sources below 300~ {\rm $\mu$ Jy}:
\begin{equation}
{\rm log} \left ( S^{2.5} \frac{dN}{dS} \right )~=~\begin{cases} 
                   \sum_{j=0}^6 a_j \left[ {\rm log} \left( S \right) \right]^j & \text{if } S \geq 300~ {\rm \mu Jy}\\
                   \sum_{j=0}^6 a_j \left[ {\rm log} \left( 0.3 \right) \right]^j & \text{if } 30 {\rm \mu Jy} \leq S \leq 300~ {\rm \mu Jy}\\
                   \sum_{j=0}^6 a_j \left[ {\rm log} \left( 10 S \right) \right]^j & \text{if } S < 30~ {\rm \mu Jy}
                   \end{cases}
\label{eqdnds1}
\end{equation}
where the flux density S is in mJy, and $a_0$ = 0.859, $a_1$ = 0.508, $a_2$ = 0.376, $a_3$ = $-$0.049, $a_4$ = $-$0.121, $a_5$ = 0.057, $a_6$  = $-$0.008.

The distribution of sources given by eqn.~\ref{eqdnds1} is shown in Fig.~\ref{fig:dnds}.
Given the angular size of the H{\sc i} source, using eqn.~\ref{eqdnds1} we can determine  the number distribution of radio continuum sources as a function of spectral flux density that would overlap with the H{\sc i} source on the sky.
The NRAO VLA Sky Survey (NVSS) and Sydney University Molonglo Sky Survey (SUMSS) are radio continuum surveys at 1.4 GHz ($\sim$ 45 \arcs~resolution), which in between them cover the entire sky and are complete to $<$10 mJy \citep{1998AJ....115.1693C,2003MNRAS.342.1117M}.
Data from these surveys would be used to identify the bright radio continuum sources in future H{\sc i} IM experiments, which will then be masked out from the images (this would be done before `foreground subtraction' of radio continuum sources).
Therefore we set a conservative upper limit of 100\,mJy for radio continuum sources which will contribute to the absorption that we aimed to quantify.
In reality the H{\sc i} flux absorbed will be slightly less than what we calculate here as radio continuum sources with lower flux densities will also be masked out given the completeness limits of NVSS and SUMSS.
But the reduction in absorption will be small, as sources with flux densities between 10\,mJy and 100m\,Jy do not contribute significantly to the total absorption (see Fig.~\ref{fig:vsmax}).

\subsection{Distribution of background sources with redshift}
\label{ssec:fz}

Of the radio continuum sources, only those that are at a redshift higher than the H{\sc i} source redshift will contribute to the H{\sc i} absorption.
Therefore in order to determine the number of radio continuum sources in the background of the voxel of sky being observed, we require the redshift distribution of radio continuum sources.
We used the redshift distribution of radio continuum sources from the simulated radio continuum sky of \citet{2008MNRAS.388.1335W}.
We needed to use results from simulations because even the deepest observational studies of the redshift distribution of radio continuum sources are complete only to 10\,mJy in flux density \citep[see e.g.][]{2008MNRAS.385.1297B}.
In Section~\ref{sec:res} we discuss how our final results are affected if the redshift distribution of sources varies as compared to what we assumed here.

\citet{2008MNRAS.388.1335W} provides a catalogue of $\sim$320 million simulated radio sources down to a flux density limit of 10 nJy at five different frequencies.
We used the catalogued values for 1.4 GHz.
The simulation uses a semi-empirical approach designed with SKA goals of galaxy evolution and large scale structure in mind.
Sources are drawn from observed (extrapolated when needed) luminosity functions placed on top of an underlying dark matter density field with biases which reflect the measured large-scale clustering of radio sources.
The radio sources are of five distinct types: radio loud and radio quiet active galactic nuclei of both type I and type II Fanaroff-Riley structural classes, as well as star-forming galaxies which in turn are split into quiescent and starbursting galaxies.
We determined the redshift distributions of the radio sources from the catalogue after binning them in flux density.
The normalized versions of these redshift distributions are shown in Fig.~\ref{fig:contz}.
It is of interest to note that for radio sources with flux density below 1 mJy, the distribution starts to have a more pronounced tail at high redshifts.
We used these normalized distributions to calculate the fraction, $f_{z,HI}$, of the total number of radio continuum sources that would lie behind our H{\sc i} source of interest, based on its redshift.

\subsection{Covering fraction for background sources}
\label{ssec:fc}

We also needed to calculate what fraction a background source would be incident on the H{\sc i} source, given that some of the background sources may be larger in size than the H{\sc i} source.
We used the relation between median angular sizes of radio sources at 1.4 GHz and their flux densities found by \citet{1990ASPC...10..389W} for the flux density range between 0.1 mJy and 10 Jy:
\begin{equation}
d_{med}~=~2^{\prime \prime} S^{0.3},
\label{eqss}
\end{equation}
to define a flux density $S_{eq}$ for which the background source's angular size ($\Omega_b$ in sr) matches that of the H{\sc i} source ($\Omega_{HI}$ in sr).
More recent deep radio continuum surveys at 1.4 GHz find results consistent with the above relation \citep[e.g.][]{2003A&A...403..857B,2018MNRAS.481.4548P}.
Thus the incident flux density on the H{\sc i} source from a background source with flux density $S$ is $f_c S$, where
\begin{equation}
f_c~=~\begin{cases}
              1 & \text{if } S < S_{eq} \\
              \frac{\Omega_{HI}}{\Omega_b} & \text{if } S \geq S_{eq}.
              \end{cases}
\end{equation}
We can expect radio source sizes to vary with redshift, and thus the distribution of source sizes is in principle coupled to the redshift distribution of sources (Fig.~\ref{fig:contz}).
\citet{1990ASPC...10..389W} considered all sources from multiple surveys irrespective of redshift, and thus any such coupling has in principle been incorporated in eqn.~\ref{eqss}.

We note though that the way we take covering fraction into account is an over-simplification, born out of necessity as there is very little information to fall back upon.
The limited information on the variation of radio continuum source sizes compared to H{\sc i} source sizes, especially beyond the very local Universe where the H{\sc i} distribution can be mapped in emission, also restricts us from quantitatively determining the error on our final results due to our assumed covering fractions.
This information, like many others discussed in Section~\ref{sec:res}, needs to come from future deep surveys radio continuum and H{\sc i} in emission by the SKA.

\subsection{Optical depth for a certain H{\sc i} column density}
\label{ssec:od}

\begin{table}[t!]
\begin{center}
\caption{Sample of DLAs and sub-DLAs used to determine the integrated optical depth -- column density relation.}
\label{tab:dla}
\begin{tabular}{cccc}
\hline
QSO&$z_{abs}$ & $\tau$dv&Reference$^*$\\
\hline
  0738+313&0.0912  &measured &1 \\
  0738+313&0.2212  &measured &1 \\
  0952+179&0.2378  &measured &1 \\
  1127-145&0.3127  &measured &1 \\
  1229-021&0.3949  &measured &1 \\
  0235+164&0.5241  &measured &1 \\
  0827+243&0.5247  &measured &1 \\
J1431+3952&0.6039  &measured &1 \\
  1331+305&0.6921  &measured &1 \\
  2355-106&1.1727  &measured &1 \\
J1623+0718&1.3367  &measured &1 \\
  2003-025&1.4106  &measured &1 \\
  1331+170&1.7763  &measured &1 \\
  1157+014&1.9436  &measured &1 \\
  1755+578&1.9698  &measured &2 \\
  1850+402&1.9888  &measured &2 \\
  0458-020&2.0394  &measured &1 \\
  2039+187&2.1920  &measured &1 \\
  0311+430&2.2890  &measured &1 \\
  0438-436&2.3474  &measured &1 \\
\hline                        
  1122-168&0.6819  &limit    &1 \\
  0454+039&0.8596  &limit    &1 \\
  2149+212&0.9115  &limit    &1 \\
J0407-4409&1.913   &limit    &2 \\
  1230-101&1.931   &limit    &2 \\
  0347-211&1.9470  &limit    &1 \\
  1452+502&1.969   &limit    &2 \\
J0733+2721&1.9758  &limit    &2 \\
  1215+333&1.999   &limit    &2 \\
  0620+389&2.031   &limit    &1 \\
J1412+1257&2.0632  &limit    &2 \\
J0845+4257&2.0652  &limit    &2 \\
J1634+3203&2.0923  &limit    &2 \\
J0214+0632&2.1075  &limit    &2 \\
J1419+0603&2.1080  &limit    &2 \\
  1645+635&2.1253  &limit    &2 \\
  0149+335&2.141   &limit    &1 \\
  0528-250&2.141   &limit    &2 \\
J1301+1904&2.1482  &limit    &2 \\
J1522+2119&2.1709  &limit    &2 \\
  1228-113&2.193   &limit    &2 \\
J1224+1947&2.2104  &limit    &2 \\
J0934+3050&2.2143  &limit    &2 \\
J1529+1904&2.220   &limit    &2 \\
  1048+347&2.2410  &limit    &1 \\
J1138+0428&2.2427  &limit    &2 \\
  0432-440&2.297   &limit    &2 \\
J1138+0428&2.3292  &limit    &2 \\
J0912+4126&2.3867  &limit    &2 \\
J0934+3050&2.3883  &limit    &2 \\
  0201+365&2.462   &limit    &2 \\
J0214+0157&2.4886  &limit    &2 \\
J1406+3433&2.4989  &limit    &1 \\
\hline                             
\end{tabular}
\end{center}
\flushleft{$^*$: 1 -- \citet{2014MNRAS.438.2131K}, 2 -- private communication.}
\end{table}

\begin{figure}
\begin{center}
\includegraphics[width=3.7truein]{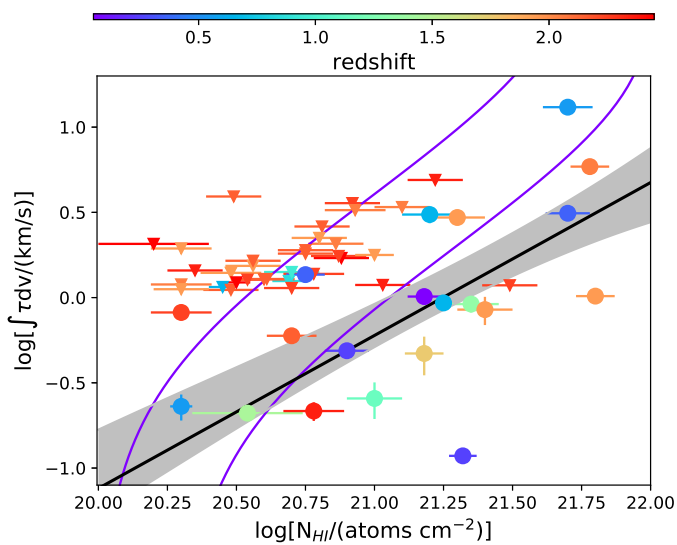} 
\end{center}
\caption{The variation of the measured integrated optical depth with the measured H{\sc i} column density  in DLAs and sub-DLAs up to a redshift of $z=2.5$. The points are colour-coded according to their respective redshifts following the colourbar on top. {\it Filled circles} represent data with measured optical depths, whereas {\it downward pointing triangles} represent data with upper limits on optical depth. The {\it black bold line} represents the linear fit to the censored data, with the {\it grey shaded region} representing the uncertainty on the fit determined using bootstrapping datasets (see Section~\ref{ssec:od} for details). The {\it violet lines} encompass the region occupied by observations of various sightlines within the Galaxy ($z=0$) with both H{\sc i} absorption and emission measurements \citep{2011ApJ...737L..33K}.}
\label{fig:tau}
\end{figure}

The amount of flux absorbed is proportional to the optical depth offered by the H{\sc i} column density of the absorber.
Specifically, the quantity of interest is the `integrated' optical depth over the full width of the absorption.
This relation between integrated optical depth and column density is at the heart of our calculations, and therefore we needed to base it on observations as opposed to expectations based on theoretical calculations or simulations.
Empirically this relation can be quantified in H{\sc i} sources up to intermediate redshifts by using observations of Damped Lyman-$\alpha$ Systems (DLAs) which contain most of the neutral hydrogen in the Universe \citep{2005ARA&A..43..861W}.
Ultraviolet/optical spectroscopic observations through which DLAs are identified directly provide a measure of their H{\sc i} column densities, and follow-up H{\sc i} absorption observations are needed to measure their integrated optical depths. 

We compiled an up-to-date list of DLAs and sub-DLAs (systems with H{\sc i} column densities just below the DLA limit of $2 \times 10^{20}$ \acc) for which the H{\sc i} column density has been measured, and the integrated optical depth has either been measured or has a robust upper limit.
Of the ones with measured integrated optical depths, one is at a redshift of 3.39, and the rest span a redshift range of 0.09 to 2.35.
\citet{2014MNRAS.438.2131K} find statistically significant evidence for an increase in spin temperatures beyond $z\approx2.4$ by combining observed upper limits and detections of H{\sc i} absorption in of DLAs and sub-DLAs.
An increased spin temperature would imply lower integrated optical depths for the same H{\sc i} column density. 
Considering this fact combined with the paucity of measurements beyond $z>2.35$, we restricted our estimates to a redshift of $z=2.5$ in this paper.

All the DLAs and sub-DLAs from our compilation with redshift $z<2.5$ are listed in Table~\ref{tab:dla}.
Column 1 lists the quasar in whose sightline the Lyman-$\alpha$ absorption is observed, column 2 gives the redshift at which the absorption is detected, column 3 lists whether the integrated optical depth ($\tau$dv) was measured or only an upper limit was obtained using follow-up H{\sc i} absorption measurements.
Column 4 lists whether the reference from which the measured values have been taken is \citet{2014MNRAS.438.2131K} -- otherwise they were provided by Nissim Kanekar (private communication).

In Fig.~\ref{fig:tau} the measured integrated optical depths and upper limits of the DLAs and sub-DLAs mentioned above are plotted against their column densities.
From the figure we can see that for similar H{\sc i} column densities, on average DLAs have lower integrated optical depths compared to the values in our own Galaxy.
We can also see that there is no obvious trend with redshift for the relationship between integrated optical depth and H{\sc i} column density.
We fitted the following relation using the Akritas-Theil-Sen estimator for censored data \citep{aks95}, with the Turnbull estimate of intercept, using the {\sc cenken} function in {\sc R} programming language:
\begin{equation}
{\rm log} \left ( \frac{\int \tau dv}{\rm km s^{-1}} \right ) = 0.9~{\rm log} \left ( \frac{N_{HI}}{\rm atoms~cm^{-2}} \right ) - 19.1,
\label{eqtau}
\end{equation}
plotted as the bold black line in Fig.~\ref{fig:tau}.
In order to estimate the uncertainty on the above relation, we bootstraped the data to create multiple datasets by varying each measurement within the 1$\sigma$ error on either side of the respective measurement (upper limits remain unchanged).
The fits to the bootstrapped censored datasets is shown as the grey shaded region in Fig.~\ref{fig:tau}.
We used the above relation to arrive at our main results, and the uncertainty on the fitted relation to estimate one of the major uncertainties in our result.


\subsection{Total flux emitted and absorbed by a H{\sc i} source}

Finally, we needed to combine the various steps listed above to estimate the total flux absorbed within a given voxel, and compare it to the total flux emitted from the same voxel.
We used the catalogued properties for H{\sc i} sources from the O9 catalogue to calculate the emitted flux $F_{emit}$.
We also used the assumption that the emission is spread uniformly across the velocity width.
We note that column definitions from the O9 catalogue are quoted hereon when describing parameters used in our calculations.
The `velocity-integrated line flux of the H{\sc i} line' (in Jy \kms) of a particular H{\sc i} source is divided by the source's H{\sc i} velocity width as defined by the `line width at 20\% of peak luminosity density (already corrected for the galaxy inclination)' (in \kms).
This value is then multiplied by  $\Delta v$ to obtain $F_{emit}$.
$\Delta v$ (in \kms) is equal to the observation channel width $\Delta v_{cw}$, when $\Delta v_{cw}~<~\Delta v_{HI}$, and $\Delta v$ is equal to $\Delta v_{HI}$ when $\Delta v_{cw}~\geq~\Delta v_{HI}$.

To calculate the flux absorbed by each such source in the O9 catalogue, we started with the catalogued value for `H{\sc i} radius where $\Sigma_{HI}$ is at 10\% of its maximum value' (in arcseconds) to be the semi-major axis of the elliptical disk.
We used the catalogued value for the `minor axis/major axis for H{\sc i}' to calculate the size of the semi-minor axis of the elliptical disk.
The major and minor axis values in combination with the angular diameter distance determined from the catalogued values of `comoving distance to the object' (in Mpc) and `apparent redshift (including Doppler correction)', is used to calculate the area of the disk.
The average H{\sc i} column density over the disk for a given source is calculated by dividing the `HI mass' (in $M_{\odot}$) of the source by the area of its elliptical disk in $pc^2$.
This is based on the fact that given the low spatial resolution of upcoming H{\sc i} IM experiments, they will only be sensitive to the average column density over the disk of the H{\sc i} source.

Thereafter we used the formalism developed in the previous sections.
Given the average column density $N_{HI}$ for the source, we calculated the $\int \tau (v) dv$ using eqn.~\ref{eqtau}.
Next, we drew upon our assumption from Section~\ref{ssec:his} that the H{\sc i} is distributed uniformly over some velocity width $\Delta v_{HI}$, and divided $\int \tau (v) dv$ by $\Delta v_{HI}$ to determine an effective optical depth per unit velocity $\tau_{e}$.
We used the catalogued value of `line width at 20\% of peak luminosity density (already corrected for the galaxy inclination)' (in \kms) as the H{\sc i} velocity width $\Delta v_{HI}$.

If the total flux density incident on the H{\sc i} source is $S_0$, the total reduction in flux due to H{\sc i} absorption produced by the H{\sc i} source in consideration is $\int S_0 (1-e^{- \tau_{e}}) dv~=~S_0 (1-e^{- \tau_{e}}) \Delta v$, where $\Delta v$ is defined in the same way as for the case of emission.
In order to determine $S_0$ we summed over all the background sources whose flux would be absorbed by the H{\sc i} source under consideration.
We therefore obtained the total flux absorbed by the H{\sc i} source from the incident flux of radio continuum sources behind it by
\begin{equation}
F_{abs}~=~f_{z,HI} \big[ \int_{S_{min}}^{S_{max}} f_c \frac{dN}{dS} \Omega_{HI} dS \big] (1-e^{- \tau_{e}}) \Delta v,
\label{eqsa}
\end{equation}
where $\frac{dN}{dS}$, $f_{z,HI}$ and $f_c$ have been defined in Sections~\ref{ssec:dnds}, \ref{ssec:fz} and \ref{ssec:fc} above.
The observed number of radio sources at 1.4 GHz falls sharply below 1\,$\mu$Jy in the model we use (there are no actual observations of radio continuum sources with flux densities below 1\,$\mu$Jy), and therefore we chose $S_{min}~=~10^{-7}$~Jy.

We repeated the above exercise for all sources within a voxel.
Thus we arrived at the quantity of interest: the ratio of the total absorbed flux to the total emitted flux for all sources within the voxel, $\sum (F_{abs}/F_{emit})$.

\subsection{Summing over sources within a voxel}

\begin{table}
\begin{center}
\caption{Size of O9 simulation box width at various redshifts}
\label{tab:cov}
\begin{tabular}{ccc}
\hline
Redshift&Simulation box side&No. of sources for\\
& (degrees) &1 MHz wide channel\\
\hline
0.1&116.5&2.9$\times$10$^5$\\
0.5& 21.4&3.0$\times$10$^5$\\
1.5&  9.0&4.8$\times$10$^5$\\
2.5&  6.7&5.4$\times$10$^5$\\
\hline
\end{tabular}
\end{center}
\end{table}

The ratio of the total absorbed flux to the total emitted flux for all H{\sc i} sources within a given voxel is what we wanted to calculate.
In the O9 simulated sky, the full simulated sky area at any redshift is a square of side $\sim$500$h^{-1}$~Mpc.
Over the large redshift range we are interested in for this work, the comoving size of the simulated sky area of the O9 simulations being constant results in a large variation in the angular size of the simulation boxes, as can be seen in Table~\ref{tab:cov}.
At each redshift we focused on voxels defined by equally wide channel widths in frequency, between $0.1$ and $10$~MHz, which cover the likely range in frequency widths to be used by upcoming H{\sc i} IM surveys.
Given that the same width in frequency units translate to increasingly large velocity widths with increasing redshift, we find from Table~\ref{tab:cov} that the total number of sources actually increases (within a factor of 2) with redshift.
Within the channel width range of $0.1$ and $10$~MHz, the total number of sources vary between a few times $10^4$ to a few times $10^6$.
Thus for all redshifts and all channel widths we integrated over statistically large number of sources thereby determining true `average' values for the calculated quantities.

For each redshift and channel width combination, those sources from the O9 catalogue are considered which have some part of their H{\sc i} emission lying within the channel width.
Such sources are found by considering the catalogued values of `apparent redshift (including Doppler correction)' in combination with their H{\sc i} velocity widths.
The physical dimensions of each of the H{\sc i} sources are calculated from their sizes in arcseconds using angular diameter distances for each source, which in turn are calculated using their catalogued values for `comoving distance' and `apparent redshift'. 

\section{Results and discussion}
\label{sec:res}

\subsection{Variation of absorbed-to-emitted flux with redshift}

\begin{figure*}
\begin{center}
\begin{tabular}{cc}
{\mbox{\includegraphics[width=3.5truein]{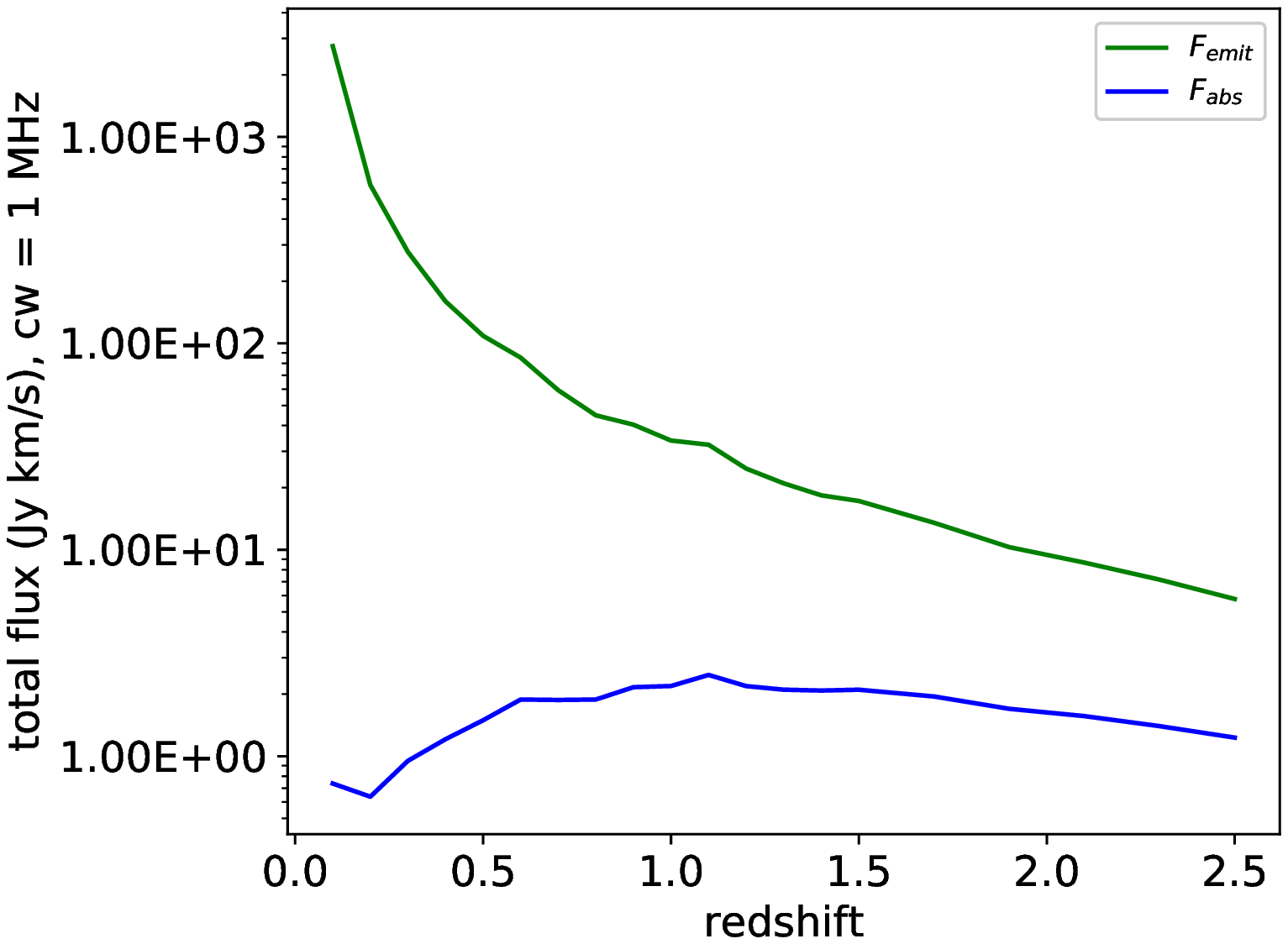}}}&
{\mbox{\includegraphics[width=3.5truein]{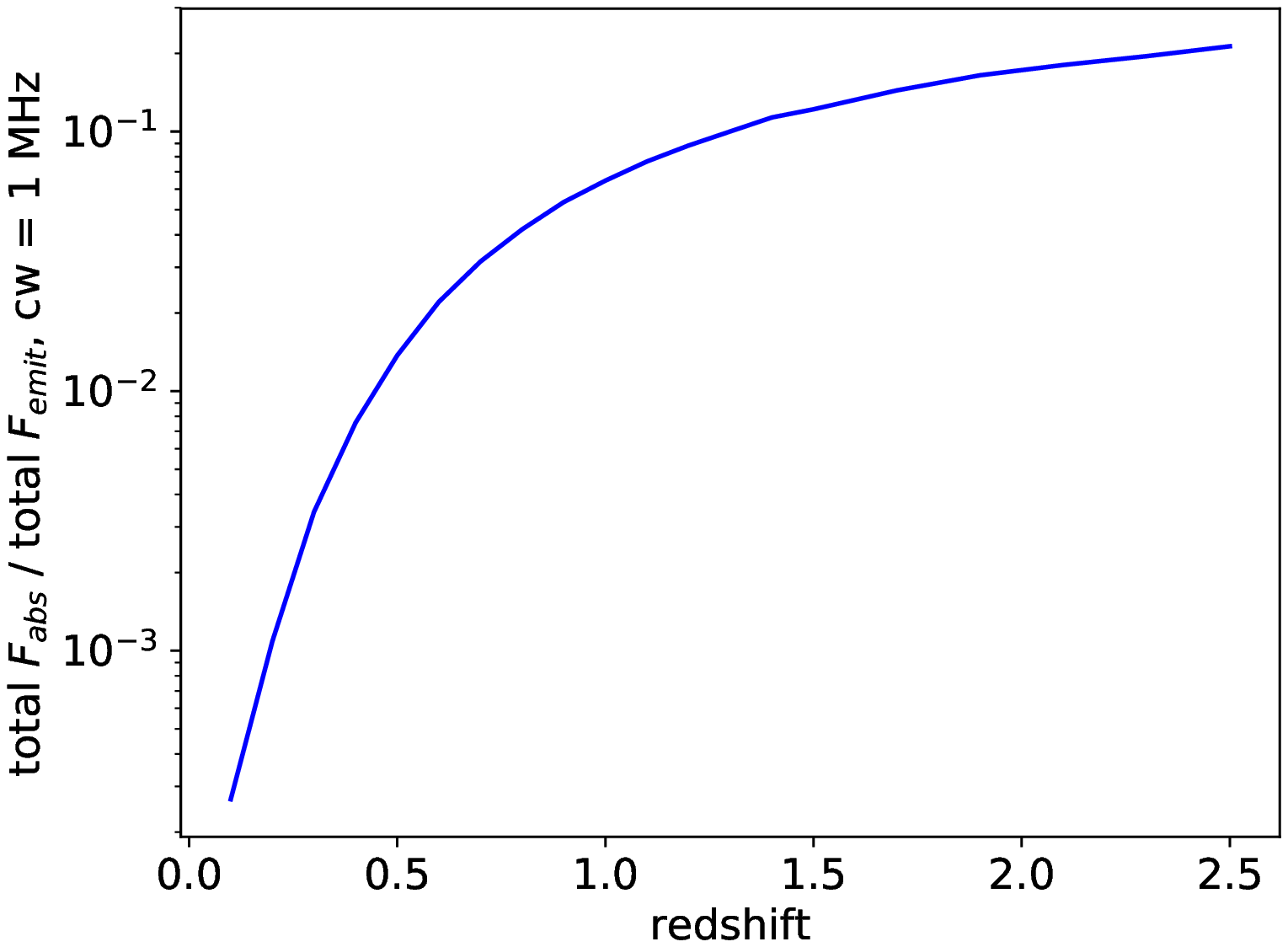}}}\\
\end{tabular}
\end{center}
\caption{The variation of total absorbed and emitted fluxes ({\it left panel}) and the ratio of the total absorbed-to-emitted flux ({\it right panel}) using the O9 simulated sky and the assumptions listed in Section~\ref{sec:cal}. The results are for all sources within voxels defined by channel width of 1 MHz, placed at various redshifts. {Note though that the ratio of the total absorbed-to-emitted flux remains unchanged for channel widths ranging between 0.1 and 10 MHz.}}
\label{fig:res}
\end{figure*}

{Here we present the results of our estimates for all H{\sc i} sources at a particular redshift within a voxel defined by the O9 simulation box size of $\sim$500$h^{-1}$~Mpc and a representative constant channel width of 1 MHz.}
We find the H{\sc i} emission and absorption fluxes increase with increasing channel width as the number of sources within a voxel increases, for channel widths between $0.1$ and $10$~MHz. 
What we also find as a consequence of always sampling a large number of sources for our chosen range of channel widths, is that the H{\sc i} emission and absorption fluxes show the exact same variation with varying channel width.
As we are interested in the relative variation between H{\sc i} emission and absorption, hereafter we report the results only for the representative channel width.

{In Fig.~\ref{fig:res} we plot for all H{\sc i} sources within each voxel, the variation with redshift of the total emitted flux, the total absorbed flux, and the ratio of the total absorbed flux to total emitted flux.
The total H{\sc i} emission flux in Fig.~\ref{fig:res} shows a marked decline with increasing redshift.
The decrease is mainly driven by the increasing luminosity distance, as the comoving volume of sky remains constant with redshift for our calculations.
The actual variation though of the total emitted flux with redshift is driven by a complex combination of factors like source numbers varying with redshift (Table~\ref{tab:cov}), and the nature of the sources themselves changing with redshift following the O9 simulation.
The variation of total H{\sc i} absorption flux with redshift as shown in Fig.~\ref{fig:res} is even more complex given all the factors discussed in Section~\ref{sec:cal}.}
What is worth noting is that the total H{\sc i} absorption flux actually increases between $z=0$ to $z=1.1$, and has a shallow peak followed by a slow decline.
Therefore while the total emitted flux reduces by more than two orders of magnitude between $z=0$ and $z=2.5$, the total absorbed flux varies very little over the same range of redshifts.

As can be seen from the right panel of Fig.~\ref{fig:res}, the differing variations of the total H{\sc i} emission and absorption fluxes with redshift results in the ratio of total $F_{abs}/F_{emit}$ rising steeply from $0.027$\% at $z=0.1$,  $1.4$\% at $z=0.5$, to $6.5$\% at $z=1.0$ and continuing to increase to $21$\% by $z=2.5$.
It should be noted that if instead we define voxels in terms of unit solid angle / survey beam size combined with a constant channel width, the variation of the emission and absorption fluxes will be different to that shown in the left panel of Fig.~\ref{fig:res}, but the ratio of the two will vary in the same way as shown in the right panel Fig.~\ref{fig:res}.
Our results are indicating that H{\sc i} 21\,cm absorption of flux incident from background radio continuum sources can be a major source of uncertainty in H{\sc i} IM experiments, particularly those probing higher redshifts.
These results should be taken to be indicative only, as there are large uncertainties on our final results and how much of an effect absorption will have on H{\sc i} IM experiments is very difficult to quantify precisely.
The uncertainties are a result of our limited knowledge of both the H{\sc i} sources and the radio continuum sources at intermediate to high redshifts, which forces us to use various assumptions and relations based on predictions from simulations.
Below we try to quantify the uncertainties that are present in on our final results presented in Fig.~\ref{fig:res} due the intrinsic uncertainties of the various assumptions and relations we have used.

{As stated in Section~\ref{ssec:his}, H{\sc i} IM experiments will be done at low angular resolutions given the scales of cosmological interest which are much larger than the sizes of the absorbers -- the H{\sc i} sources.
We note therefore that the power spectrum of the absorption will be approximately flat over the scales of interest for H{\sc i} IM.}

\subsection{Background source flux density range which affects the results most}
 
\begin{figure}
\begin{center}
\includegraphics[width=3.5truein]{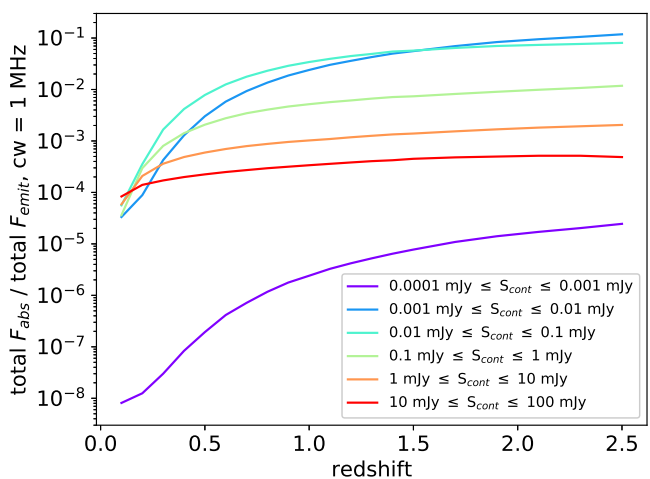} 
\end{center}
\caption{The ratio of the total absorbed to total emitted fluxes for all sources within voxels defined by channel width of 1 MHz placed at various redshifts, for various ranges of background radio continuum source fluxes.}
\label{fig:vsmax}
\end{figure}

The results in this study are for a maximum radio continuum source flux density of 100 mJy on the assumption that stronger sources are cataloged and will be removed. 
In order to understand which flux range of background sources affects the results most, we repeated the study for the chosen channel width of 1 MHz and with varying ranges of radio continuum source flux density.
The variation of total $F_{abs}/F_{emit}$ with redshift for varying ranges of background radio continuum source flux densities is presented in Fig.~\ref{fig:vsmax}.
We find that the major contribution to the H{\sc i} absorption is from the flux incident from background sources with flux densities in the range $1 \mu$Jy $\leq S_{cont} \leq 0.1$~mJy, that is around the peak of the number distribution of sources when plotted against flux density as in the right panel of Fig.~\ref{fig:dnds}.
We can now understand that the variation of the total absorbed flux with redshift in the left panel of Fig.~\ref{fig:res} is driven by the the cumulative number of radio continuum sources in the flux density range $1 \mu$Jy $\leq S_{cont} \leq 0.1$~mJy behind a voxel placed at a particular redshift (which can be calculated from the distributions shown in the lower panel of Fig.~\ref{fig:dnds}).

\subsection{Effect of uncertainty in sub-mJy number counts of background sources}

\begin{figure*}
\begin{center}
\begin{tabular}{cc}
{\mbox{\includegraphics[width=3.5truein]{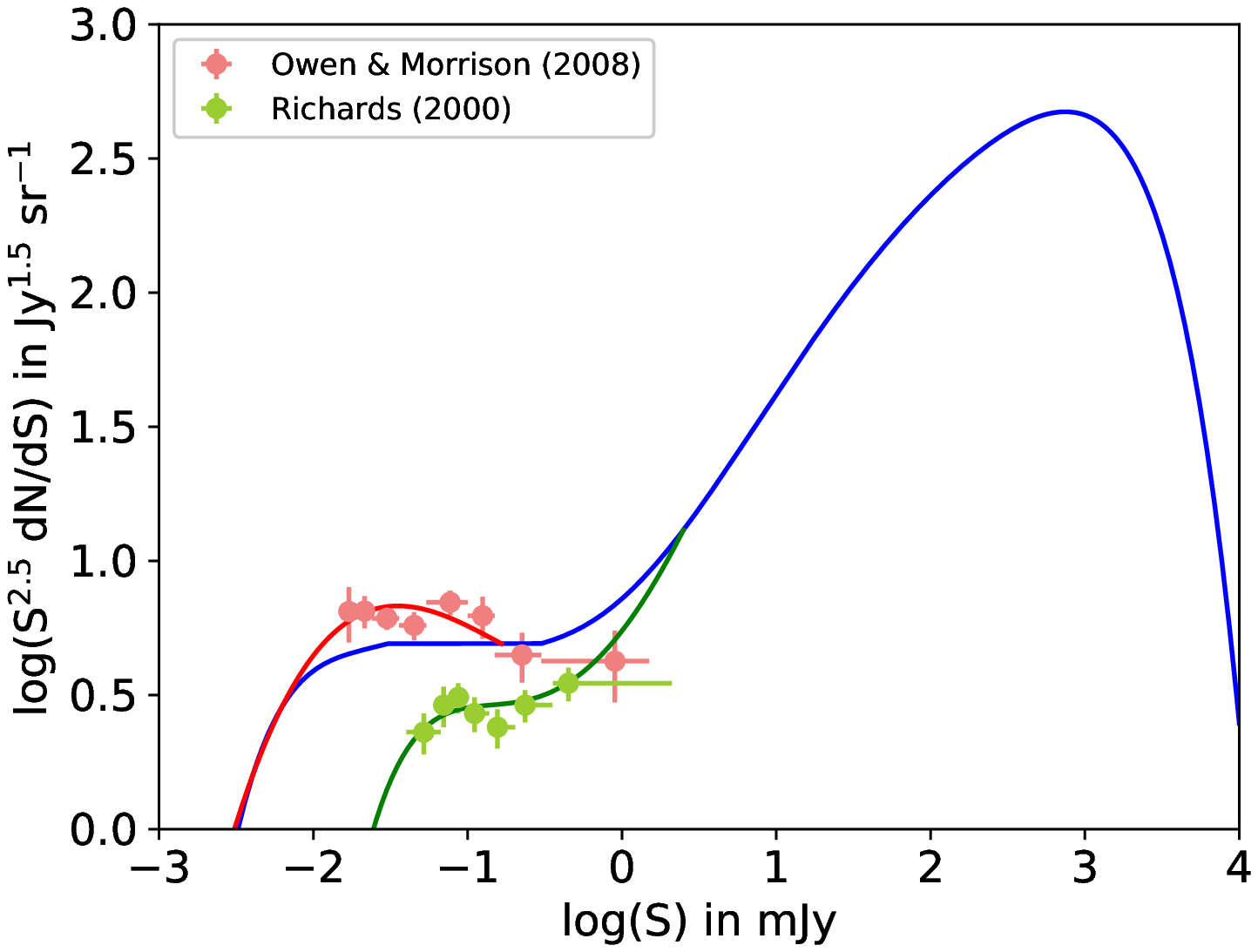}}}&
{\mbox{\includegraphics[width=3.5truein]{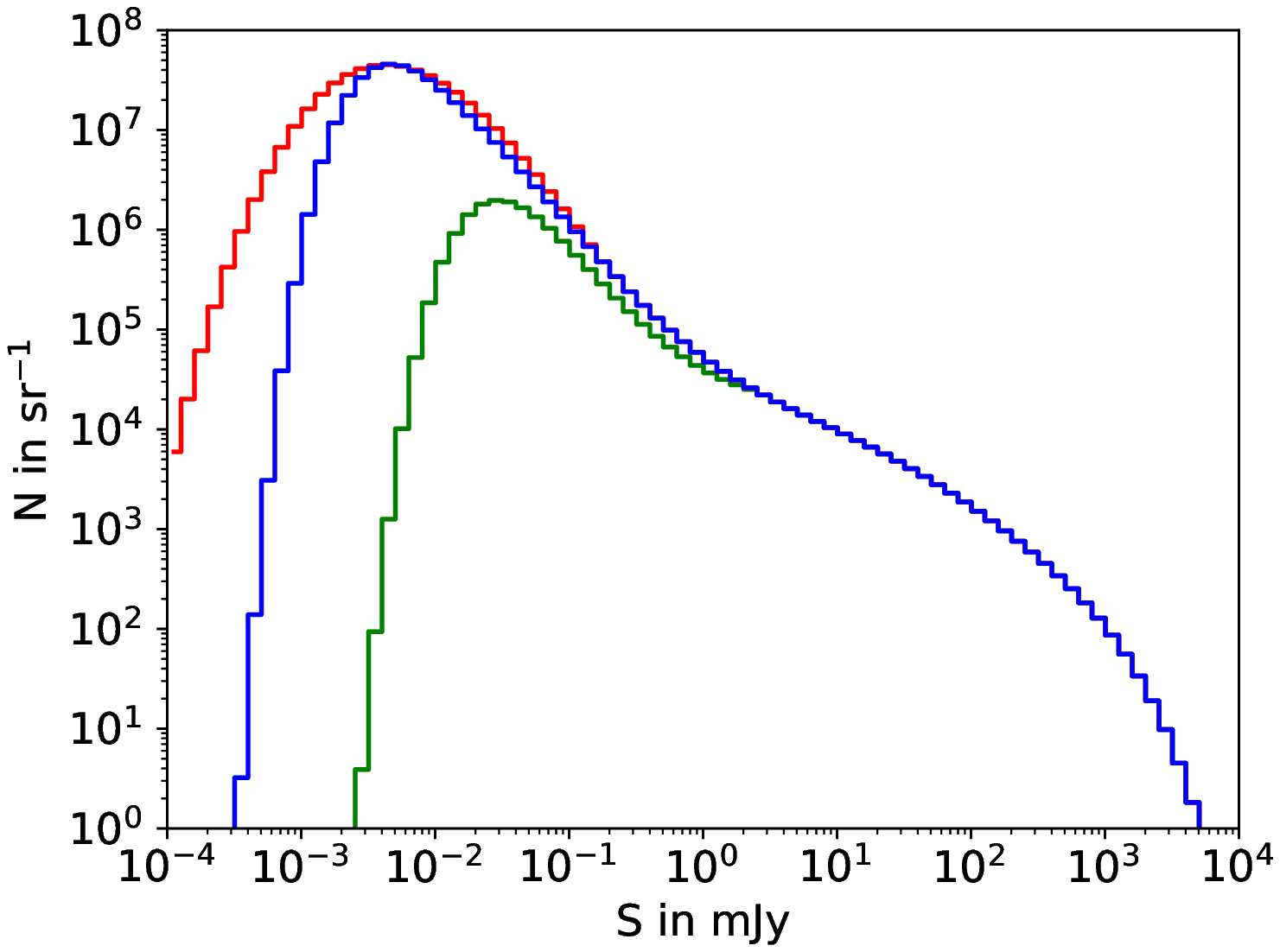}}}\\
\end{tabular}
\end{center}
\caption{The varying number counts of radio continuum sources at 1.4 GHz for sub-mJy flux densities. Euclidean normalized differential ({\it left panel}) and total ({\it right panel}) as functions of flux density. The red and green points with error bars in the left panel are measurements from \citet{2008AJ....136.1889O} and \citet{2000ApJ...533..611R} respectively. The blue curves in both panels is the fit from \citet{2014MNRAS.441.2555H} that we use to calculate our results, whereas the red and green curves are our fit to the red and green points in the left panel. Our fits are extensions of the \citet{2014MNRAS.441.2555H} fit at the low flux ranges where the red and green points deviate from the blue curve (see text for details). We note that in the right panel the x-axis starts an order of magnitude lower as compared to the left panel, in order to accentuate the variation in the number count at the low flux end from different measurements of the same.}
\label{fig:dndsa}
\end{figure*}

\begin{figure}
\begin{center}
\includegraphics[width=3.5truein]{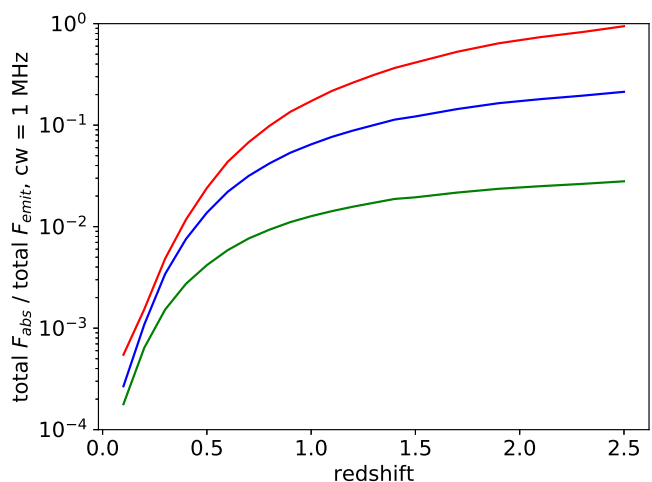} 
\end{center}
\caption{{The ratio of the total absorbed to total emitted fluxes for all sources within voxels defined by channel width of 1 MHz placed at various redshifts using the O9 simulated sky, for varying sub-mJy radio continuum source number counts. The {\it solid blue} line is identical to the result presented in the right panel of Fig.~\ref{fig:res}. The {\it solid red and green} lines show the results for the extreme version of the number counts as represented by \citet{2008AJ....136.1889O} counts and \citet{2000ApJ...533..611R} counts respectively, which are shown in Fig.~\ref{fig:dndsa}. }}
\label{fig:vns}
\end{figure}

The number counts of radio continuum sources at sub-mJy flux densities remains an open area of research.
As described in Section~\ref{ssec:dnds}, we have used the relation from \citet{2014MNRAS.441.2555H} which models the effect of an exaggerated population of faint sources below 0.3 mJy at 1.4 GHz.
\citet{2014MNRAS.440.3113H} shows that eqn.~\ref{eqdnds1} reproduces the average low ($<$0.1 mJy) flux density source counts observed by deep surveys of the radio sky at 1.4 GHz \citep{2006MNRAS.371..963B,2006ApJS..167..103F,2008AJ....136.1889O,2009MNRAS.397..281I,2010ApJS..188..178M,2011ApJ...740...20P}.
The field-to-field scatter in number counts from these and other determinations of the number counts at sub-mJy flux densities \citep[e.g.][]{2000ApJ...533..611R,2008ApJ...681.1129B,2008MNRAS.386.1695S} is large and cannot be simply explained away by sample variance \citep{2013MNRAS.432.2625H}.
It has been suggested though that the observed scatter is a result of data processing differences and calibration errors \citep[e.g. see discussion in][]{2009MNRAS.397..281I,2014MNRAS.441.2555H}.
How the number counts might look if data processing and calibration errors are accounted for is beyond the scope of this work.
In order to understand the effect this scatter can have on our results, we simply take into account all published number counts and consider the number counts from two studies which envelop the observed scatter in sub-mJy number counts, \citet{2008AJ....136.1889O} and \citet{2000ApJ...533..611R}.

In the left panel of Fig.~\ref{fig:dndsa} we plot the observed Euclidean normalized number counts from these two studies.
We create two alternate distributions of radio source counts against flux density by deviating from the \citet{2014MNRAS.441.2555H} fit at the low flux density end using fits to the data points from \citet{2008AJ....136.1889O} and \citet{2000ApJ...533..611R}, shown as red and green curves respectively in the left panel of Fig.~\ref{fig:dndsa}.
The right panel of Fig.~\ref{fig:dndsa} shows how the actual distribution of sources with flux density varies at the low flux density end for these fits, as compared to that given by the fits from \citet{2014MNRAS.441.2555H}.
What is apparent from this panel is that even a small variation in the Euclidean normalized distribution at low flux densities has a very large impact on the total number of low flux density sources.
This is reflected in how much $F_{abs}/F_{emit}$ varies depending on the distribution of radio source counts we choose, as shown in Fig~\ref{fig:vns}.
When considering the radio source count distribution which is consistent with the data from \citet{2008AJ....136.1889O}, total $F_{abs}/F_{emit}$ is $0.055$\% at $z=0.1$,  $2.4$\% at $z=0.5$, $17$\% at $z=1.0$, and $94$\% by $z=2.5$.
In contrast, when considering the radio source count distribution which is consistent with the data from \citet{2000ApJ...533..611R}, total $F_{abs}/F_{emit}$ is $0.018$\% at $z=0.1$,  $0.4$\% at $z=0.5$, $1.2$\% at $z=1.0$, and $2.8$\% by $z=2.5$.
Comparing with our main result as shown in Fig~\ref{fig:res}, we find that the scatter in the number counts of sub-mJy radio continuum sources among various measurements can vary our results by almost an order of magnitude in either direction.
As the number counts for sub-mJy flux densities are uncertain, deeper radio continuum surveys are very much necessary to constrain the number counts and thus the effect of absorption on H{\sc i} IM experiments.
One important additional caveat is that since the low flux density sources are mostly star forming (dwarf) galaxies which are clustered, the absorption of the H{\sc i} signal will be scale-dependent and can create a scale-dependent bias in the final results of an H{\sc i} IM experiment.

\subsection{Effect of uncertainty in optical depth vs. column density relation for H{\sc i} sources}

\begin{figure}
\begin{center}
\includegraphics[width=3.5truein]{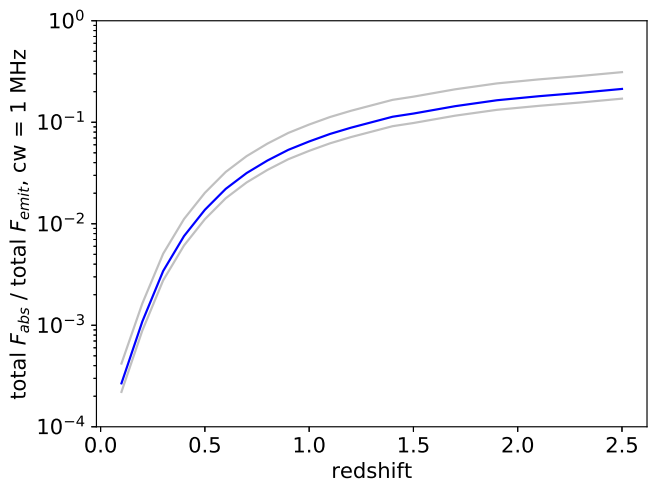} 
\end{center}
\caption{{The  {\it solid blue} line ratio of the total absorbed to total emitted fluxes for all sources within voxels defined by channel width of 1 MHz placed at various redshifts using the O9 simulated sky, identical to the result presented in the right panel of Fig.~\ref{fig:res}. The {\it grey} lines show the results due to the uncertainty of the relation between integrated optical depth and H{\sc i} column density, i.e. the grey shaded region in Fig.~\ref{fig:tau}. The scale of the plot is identical to Fig.~\ref{fig:vns}.}}
\label{fig:vtc}
\end{figure}

Another important relation defining the results presented in Fig.~\ref{fig:res} is the empirical relation between integrated optical depth and H{\sc i} column density discussed in Section~\ref{ssec:od}.
This relation is given in eqn.~\ref{eqtau}, which is based on a fit to observations with large intrinsic scatter as can be seen in Fig.~\ref{fig:tau}.
In Fig.~\ref{fig:vtc} we show how our results change on considering the error on the fit determined by bootstrapping the data, the grey shaded region in Fig.~\ref{fig:tau}.
If we first consider the upper envelope of fits to bootstrapped data,
the total $F_{abs}/F_{emit}$ goes from $0.042$\% at $z=0.1$,  $2.0$\% at $z=0.5$, to $9.5$\% at $z=1.0$ and continuing to increase to $31$\% by $z=2.5$.
On the other hand if we consider the lower envelope of fits to bootstrapped data, the total $F_{abs}/F_{emit}$ goes from $0.022$\% at $z=0.1$,  $1.1$\% at $z=0.5$, to $5.2$\% at $z=1.0$ and continuing to increase to $17$\% by $z=2.5$.
Therefore the uncertainty on the relation between integrated optical depth and H{\sc i} column density can introduce a substantial uncertainty on the quantification of then flux absorbed in an H{\sc i} IM experiment.
From Fig.~\ref{fig:tau} we can see that the point-to-point scatter around the fit between the different individual measurements is even larger than the uncertainty on the fit. 
Therefore, given the sensitivity of the results in this work to the integrated optical depth vs. H{\sc i} column density relation, we think that a detailed quantification of the relation up to intermediate redshifts is absolutely necessary for estimating the effect of absorption on H{\sc i} IM experiments.

In our calculations we assumed that the optical depth of the H{\sc i} source is distributed uniformly over the full H{\sc i} velocity width of the source under consideration, in keeping with our primary assumptions about H{\sc i} sources outlined in Section~\ref{ssec:his}.
This assumption is actually a conservative one when considering the strength of absorption, as typically the optical depth will have a non-uniform distribution with one or more peaks.
At the velocity of the peak(s) the absorbed flux would be much higher given its exponential dependence on the optical depth compared to what it would be if the optical depth is distributed uniformly over the H{\sc i} velocity width.
Therefore at high spatial resolutions, H{\sc i} absorption might start to dominate over H{\sc i} emission in an IM experiment.
In the future when the various relations affecting the quantification of the absorption strength are better constrained, we need to move beyond simply comparing the strength of the absorption to the emission as is done here.
We should aim to quantify what the power spectrum of H{\sc i} absorption itself looks like at high spatial resolutions, possibly paving the way for H{\sc i} absorption intensity mapping.

\subsection{Effect of uncertainty in redshift distribution of background sources}

\begin{figure}
\begin{center}
\includegraphics[width=3.7truein]{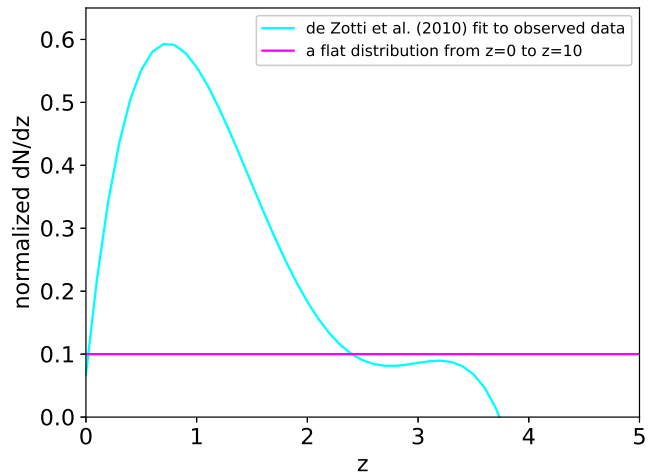}
\end{center}
\caption{Two alternate distributions of radio continuum sources with redshift {(see text for details). Compare with the original redshift distributions shown in Fig.~\ref{fig:contz}.}}
\label{fig:vnvz}
\end{figure}

\begin{figure}
\begin{center}
\includegraphics[width=3.5truein]{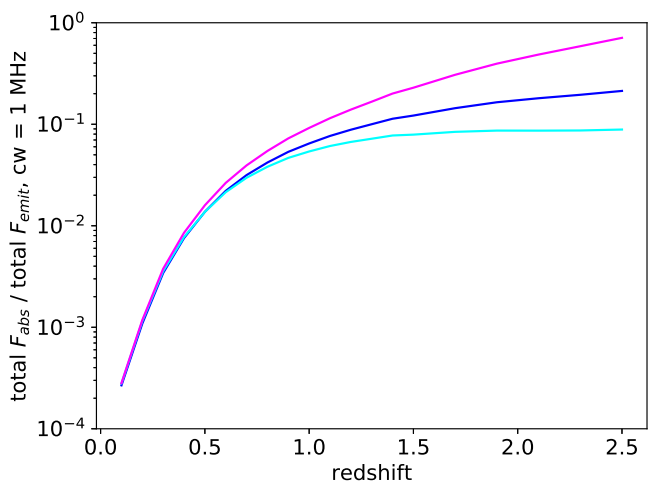} 
\end{center}
\caption{{The {\it cyan and magenta} lines show ratio of the total absorbed to total emitted fluxes for all sources within voxels defined by channel width of 1 MHz placed at various redshifts using the O9 simulated sky, for the two alternate redshift distributions of radio continuum sources shown in Fig.~\ref{fig:vnvz}. While the {\it solid blue} line shows the original result and is identical to the result presented in the right panel of Fig.~\ref{fig:res}. The scale of the plot is identical to Figs.~\ref{fig:vns} and~\ref{fig:vtc}.}}
\label{fig:vnz}
\end{figure}

In Section~\ref{ssec:fz} we describe how we use input from simulations for the distribution of radio continuum sources with redshift.
We check how our results change on using alternate distributions to those plotted in Fig.~\ref{fig:contz}.
We consider the two distributions shown in Fig.~\ref{fig:vnvz} and assume that all radio continuum sources of all fluxes follow these distributions instead of those shown in Fig.~\ref{fig:contz}. 
Alternative distribution 1 is derived from the fit to the redshift distribution from \citet{2010A&ARv..18....1D} of CENSORS sources with flux density $>$10 mJy \citep{2008MNRAS.385.1297B},
\begin{equation}
N(z)~=~1.29 + 32.37 z - 32.89z^2 + 11.13 z^3 - 1.25z^4,
\label{eqnvz}
\end{equation}
where $z$ is the redshift.
The normalized form of eqn.~\ref{eqnvz} is shown in light blue in Fig.~\ref{fig:vnvz}.
We find that this redshift distribution has some features which are in common to the simulated distribution shown in Fig.~\ref{fig:contz}: a peak at low redshifts followed by a decline. 
Unlike the redshift distributions of sources with flux densities $<100$mJy (which we use to calculate our results) shown in Fig.~\ref{fig:contz}, alternative distribution 1 has no radio sources beyond a redshift of $z \sim 3.75$.
To complement alternative 1, as alternative 2 we use an extremely hypothetical distribution where the sources are distributed uniformly up to a very high redshift -- which we take to be $z=10$.
This provides a check of how an excess of radio continuum sources at high redshifts might affect our results.

Fig.~\ref{fig:vnz} shows how the main result of the study changes on considering these alternate distributions.
We find that the redshift distribution of radio continuum sources, especially at high redshifts, can affect the amount of H{\sc i} absorption significantly at z$\gtrsim$0.5.
Any distribution which has a large fraction of the sources at higher redshift will expectedly make the absorption higher and increase the ratio of absorbed to emitted fluxes.
As alternative distribution 1 has features similar to the distributions in Fig.~\ref{fig:contz} up to a peak around $z=1$, the results on assuming alternative distribution 1 starts deviating appreciably from the results shown in Fig.~\ref{fig:res} only for $z>1$.
Even when assuming a flat redshift distribution of sources as for alternative distribution 2, the results remain unaffected for $z<0.5$, and only start to deviate appreciably from the main result around and beyond $z=1$.
Therefore it is important also to measure the redshift distribution of radio continuum sources at high redshifts, especially in the critical flux density range $0.001 mJy \leq S_{cont} \leq 0.1 mJy$ given the reasons mentioned above.
Planned SKA surveys in combination with next generation infrared space observatories should be able to determine the redshift distribution of such sources \citep{2011MNRAS.411.1547P}.

\subsection{Effects of other assumptions} 

In our study we have based the properties of all background radio continuum sources on their properties at 1.4 GHz.
Thus we have inherently assumed that the spectral index of radio sources below 1.4 GHz is on average flat.
There are observational indications though that the average radio spectral index below 1.4 GHz is $~-0.8$ \citep{2018MNRAS.474.5008D} but appears to flatten below 10\,mJy \citep[e.g.][]{2016MNRAS.463.2997M}.
The spectral indices of the sub-mJy radio continuum sources below 1.4 GHz has not been constrained observationally.
If they have steep spectral indices, our inherent assumption about a flat spectral index will turn out to be a conservative one.
The radio continuum sources will have higher flux density at the redshifted frequency of observation and the absorbed flux will also be slightly higher.
And this effect will be exacerbated with increasing redshift.
Again, future radio continuum surveys at 1.4 GHz and lower frequencies will provide the answers regarding the spectral indices of sub-mJy radio sources.

Finally, the results from this study are inherently dependent on the simulated H{\sc i} sky from the O9 simulations.
Their simulations are based on empirical recipes and physical constraints, and have been shown to match the overall observed properties of cosmic structure.
But in the end simulations are an extrapolation of existing data, and future surveys of H{\sc i} sources up to intermediate-to-high redshifts will help improve the precision of both the simulations, and predictions based on simulations like in this work.

\section{Conclusion}
\label{sec:con}

In this study we, for the first time, estimated the amount of H{\sc i} 21\,cm flux that would be absorbed by the same sources whose emission H{\sc i} IM experiments aim to detect.
We estimated the fraction of absorbed flux compared to the emitted H{\sc i} flux in voxels defined by channel widths as would be used in upcoming H{\sc i} IM experiments.
We used the H{\sc i} sky from the O9 simulations, and used multiple assumptions and relations based either on observations or input from simulations, to arrive at our result.
We find (i) generally low absorption effects at $z<0.5$, (ii) increasing absorption effects above $z=0.5$ reaching $\sim$10\% of the emission at $z=1$. 
(iii) Absorption effects continue to increase above $z=1$ and can reach 30\% of the emission signal at $z=2.5$.
The above numbers vary significantly, as large as an order of magnitude in either direction, depending on the uncertainties in the relations used to calculate the amount of absorption.
Up to z$\sim$0.5 the modelling uncertainties are not too large but above that they can become significant.
The three main relations whose impreciseness affect our results, in order of importance, are: the number counts of radio continuum sources especially at sub-mJy flux densities, the relation between integrated optical depth and H{\sc i} column density of H{\sc i} sources, and the redshift distribution of radio continuum sources up to the highest redshifts.

Therefore the results from multiple planned deep high-resolution surveys, specifically using the SKA, are required to properly quantify the effect of absorption on the H{\sc i} IM signal.
These include deep surveys of the radio continuum emission to determine the number counts, redshift distributions, and size distribution of radio sources.
SKA absorption surveys which will detect H{\sc i} absorption from multiple DLA sightlines can possibly improve the accuracy of the measured integrated optical depth vs. H{\sc i} column density relation and its variation with redshift.
SKA surveys of H{\sc i} 21\,cm emission sources up to intermediate redshifts will help fine-tune simulations like O9, and enable us to make better predictions using such simulations.
Therefore the interpretation of H{\sc i} IM surveys will require knowledge of the effects of absorption, which could bias the results.
The amount of absorption, especially at high redshifts, needs to be quantified using data from future deep surveys focusing on radio continuum, H{\sc i} absorption, and H{\sc i} emission from individual sources.

\section*{Acknowledgments}
SR would like to thank Nissim Kanekar for detailed discussions and suggestions.
We thank James Allison for helpful comments regarding material presented in Appendix~\ref{sec:app}. 
SR and CD acknowledge support from ERC Starting Grant No. 307209.
SR acknowledges that this research was partially supported by the Australian Government through the Australian Research Council funding (FT150100269).
CD is also supported by an STFC Consolidated Grant (ST/P000649/1).

\bibliographystyle{aa} 
\bibliography{sambit.bib}

\begin{appendix}

\section{Other absorptions in H{\sc i} IM}
\label{app:abs}

Here we discuss absorptions that will be imprinted in the signal picked up by an H{\sc i} IM experiment, other than  absorption of flux incident from background radio continuum sources by the H{\sc i} emitting sources themselves (the main topic of this paper).
The strength of these absorptions are likely to be insignificant compared to the main type of absorption discussed in this paper, but are listed here for completeness.

\subsection{Within the H{\sc i} emitting sources}

\subsubsection{H{\sc i} self-absorption}
\label{ssec:self}

{H{\sc i} self-absorption occurs when the H{\sc i} responsible for emission has very high optical depths ($\tau > 1$).
The self-absorption can be quite high in the highest column density regions of nearby disk galaxies \citep{2012ApJ...749...87B}.
But the volume filling factor of high column density H{\sc i} is insignificant when considering the entire disk of a gas-rich galaxy ($\sim$1\%), which is the typical H{\sc i} emitting source.
H{\sc i} IM in the foreseeable future will be done (at best) at resolutions of few arcminutes, because of which the studies will only be sensitive to the average H{\sc i} column densities over the disks of the H{\sc i} emitting sources. 
Therefore self-absorption is unlikely to be an important effect on scales relevant for H{\sc i} IM experiments, and the assumption made regarding the H{\sc i} emitting gas being optically thin is justifiable.}

\subsubsection{Absorption by electrons}
\label{ssec:sabs}

A secondary internal source of absorption of the H{\sc i} emission from a given source are electrons, the same ones which are responsible for synchrotron and free-free continuum emission from the source itself.
Such emission again has a low volume filling factor when integrated over the entire disk of a galaxy, because the radio continuum is dominated by localized dense HII regions and the electron density in such regions drops sharply with their size \citep{2009A&A...507.1327H}.

\subsection{Intervening absorption}

A set of absorptions will occur due to absorbing media situated in between the H{\sc i} emission source and the observer, including our Galaxy. 

\subsubsection{H{\sc i} absorption by other sources in voxel}

The H{\sc i} flux emitted by any individual source within a voxel being observed by an H{\sc i} IM experiment is subject to possible absorption by the H{\sc i} and the electrons responsible for radio continuum emission (see Section~\ref{ssec:sabs}) of intervening H{\sc i} sources within the voxel itself.
Based on the arguments listed in Section~\ref{ssec:sabs}, the absorption by the electrons producing the radio continuum is likely to be negligible.
As for the absorption by H{\sc i} in the intervening sources within the voxel, in addition to the low probability of H{\sc i} sources within a voxel being observed to spatially overlap each other, the optical depth responsible for the absorption will be low because of the coarse beam size.
Given the emitted H{\sc i} flux that is being absorbed is itself minuscule compared to say the radio continuum flux from background sources, such absorption is likely to be significantly sub-dominant to the absorption of flux from background radio continuum sources by the H{\sc i} sources under consideration.

\subsubsection{Absorption by electrons in intervening sources including our Galaxy}
\label{ssec:iels}

Regarding the absorption that can occur in sources between $z=0$ and the voxel being observed, there can be no absorption by the H{\sc i} in such intervening sources including the Galaxy given velocity considerations.
The velocity spread of the H{\sc i} is such that H{\sc i} IM experiments beyond the very local ($z<0.005$) Universe will remain unaffected.

There will be absorption though by the electrons in the intervening sources responsible for radio continuum emission, especially so in the Galaxy itself. 
Such absorption can occur at all frequencies below 1.4 GHz to which the H{\sc i} 21\,cm emission signal from the sources being observed might be redshifted to, even as low as $\sim$400 MHz for H{\sc i} IM experiments observing the Universe at $z=2.5$.
At these low frequencies though the amount of absorption by electrons producing synchrotron and free-free emission is likely to be small.
In intervening sources the emission will become insignificant given the small volume filling factors of the HII regions which dominates the radio continuum emission (see Section~\ref{ssec:sabs}).
As for our Galaxy, H{\sc i} IM experiments will observe the sky out of the plane of the Galaxy in order to avoid the regions with the strongest synchrotron and free-free emission.

\subsubsection{Absorption by other species}

There is also the possibility of  absorption in intervening sources and the Galaxy by other species with rest frame wavelengths which happen to match that of the redshifted H{\sc i} 21\,cm emission incident on them.
Specifically, a number of hydrogen recombination lines have frequencies in this range.
The optical depth of such absorption is negligible except at very high densities, as is found in the densest HII regions \citep{1978ARA&A..16..445B}.
Therefore as in Sections~\ref{ssec:sabs} and \ref{ssec:iels}, this type of absorption will be negligible due to volume filling factor considerations.


\subsection{Associated H{\sc i} absorption}

There is a special type of absorption which will also be included in the signal picked up by H{\sc i} IM experiments, but which does not arise along the sightline of the H{\sc i} emitting sources discussed here or in Appendix~\ref{app:abs}.
We are referring to `associated' (as opposed to `intervening') H{\sc i} absorption within radio continuum sources which happen to be in the voxel of sky being observed, by H{\sc i} clouds within the sources themselves H{\sc i} clouds \citep[see][for a review]{2018arXiv180701475M}.
H{\sc I} clouds are approximately 100 pc in size, for example see \citet{2012ApJ...749...87B}.
Therefore realistically such absorption will only be detected against sources whose radio continuum emission is dominated by a very compact core - core-dominated active galactic nuclei (AGNs) and nuclear starbursts are the obvious candidates, with edge-on viewing angles favoured.
The number of detections of associated H{\sc i} absorption are rare ($<$100 detections) even with targeted searches, especially at $z>0.25$ \citep[e.g.][]{1970ApJ...161L...9R,1986ApJ...300..190D,1989AJ.....97..708V,1998ApJ...494..175C,1999ApJ...524..684G,1999ApJ...510L..87M,2003A&A...404..871P,2003A&A...404..861V,2005A&A...444L...9M,2006MNRAS.373..972G,2011MNRAS.418.1787C,2011ApJ...742...60D,2012MNRAS.423.2601A,2013MNRAS.429.2380C,2014MNRAS.440..696A,2015A&A...575A..44G,2017A&A...604A..43M,2018MNRAS.473...59A,2018MNRAS.481.1578A,2019MNRAS.482.5597A}.
The bright continuum sources against which such absorption might be detectable will in any case be flagged in an H{\sc i} IM experiment (conversely see Appendix~\ref{sec:app} for an estimate of how many `detections' can be expected for a given H{\sc i} IM experiment when considering the brighter end of radio continuum sources).
As for the non-detectable radio continuum background, the fainter end of the radio continuum emission is increasingly dominated by star forming dwarf galaxies and the fraction of starbursts and AGNs decreases rapidly with flux \citep{2011MNRAS.411.1547P}.
Moreover given the fact that such absorption is caused by localized, the low spatial resolution of upcoming H{\sc i} experiments will dramatically reduce the amount of absorption that will be picked up.
Therefore associated H{\sc i} absorption is unlikely to be of significance for upcoming H{\sc i} IM experiments, though a detailed quantification of its contribution especially for high resolution H{\sc i} IM experiments should ideally be undertaken in the future.

\section{Detections of associated H{\sc i} absorption}
\label{sec:app}

In Section~\ref{app:abs} we mentioned how associated H{\sc i} absorption against the bright radio emission from the centres of radio continuum sources lying within the voxel of sky being observed will be picked up by H{\sc i} IM experiments.
We also reasoned why such absorption from unflagged radio continuum sources in the data is likely to be insignificant.
Here we reverse the question and ask how many associated H{\sc i} absorption detections will be possible for a given H{\sc i} IM experiment.
Thus the brightest radio continuum sources that would otherwise be flagged for H{\sc i} IM, become potential targets for new associated H{\sc i} absorption detections.
Given that detections of associated H{\sc i} absorption are limited, and H{\sc i} IM experiments will observe large areas of the sky much of which has not been explored in previous searches of associated absorption using existing facilities, data from H{\sc i} IM experiments can potentially be used to search for new associated H{\sc i} absorption detections.

\subsection{Number of detections for a test case: BINGO}

Here we calculate the potential number of detections of associated H{\sc i} absorption, given the parameters of an actual H{\sc i} IM experiment viz.
BAO from Integrated Neutral Gas Observations \citep[BINGO,][]{2016arXiv161006826B}.
Upcoming H{\sc i} IM experiments will have beam sizes between few arcminutes to tens of arcminutes.
For BINGO the beam size will be 40\arcm.
Therefore we based our calculations on the detection rates achieved by associated H{\sc i} absorption searches using single-dishes.

\begin{table*}
\begin{center}
\caption{Associated H{\sc I} absorptions that will be detectable using the H{\sc i} IM survey with BINGO covering 5000 deg$^2$ of sky.}
\label{tab:bingo}
\begin{tabular}{ccccccc}
\hline
$z_{mean}$ & 5$\sigma$ & $S_{cont}^{limit}$ & Vol. & No. of SF & No. of & total \\
~ & (mJy) & (mJy) & (Gpc$^3$) & galaxies & AGNs & detections \\
\hline
0.15 & 4.9 & 16.3 & 0.262 & 5 & 1334 & 27 \\
0.25 & 4.1 & 13.7 & 0.643 & $<$1 & 2247 & 45 \\
0.35 & 3.5 & 11.7 & 1.127 & $\ll$1 & 2598 & 52 \\
0.45 & 3.1 & 10.3 & 1.666 & $\ll$1 & 2512 & 50 \\
\hline
\end{tabular}
\end{center}
\end{table*}

\citet{2014MNRAS.440..696A} searched in archival H{\sc i} Parks All Sky Survey (HIPASS) data for associated H{\sc i} absorption within a sample of 204 radio and star-forming galaxies within a volume bounded by a maximum redshift of 0.042 and covering 34,432 deg$^2$ in the sky.
HIPASS has a rms noise of $\sim$13 mJy per 15.5\arcm~beam, with channel width of 13.2 \kms\ at z=0.
The rms noise can potentially vary significantly across HIPASS.
Therefore to ensure a $>$5$\sigma$ detection of a peak absorption of 30\%, the continuum flux density limit was set to $\sim$250 mJy and only sources above that limit (at either 843 MHz or 1.4 GHz) were targeted.
Absorption was detected in four sources, of which one is a compact starburst, two are compact active galactic nuclei (AGNs) when compared to the respective stellar disks, and the fourth one is an extended AGN.
The detection rates were 2\% for the entire sample, as well as for the star-forming and AGN dominated samples separately.
It is notable that their determinations of the total number of galaxies which are either star-forming or are AGN dominated, match quite well with the numbers predicted using the local radio luminosity functions for the two classes of galaxies from \citet{2007MNRAS.375..931M}.
\citet{2014MNRAS.440..696A} also compare their detection rates with those from the Arecibo Legacy Fast ALFA Survey (ALFALFA) pilot survey \citep{2011ApJ...742...60D}, which is a survey of 517 deg$^2$ of the sky in the redshift range 0$<$z$<$0.058 (with 10 \kms~velocity resolution and a beam size of 3.3\arcm $\times$ 3.8\arcm).
The flux density limit for continuum sources against which absorption is expected to be detected is 42 mJy, considering a 5$\sigma$ detection of at least 30\% peak absorption.
There was only a single detection of (strong) absorption, among 29 potential detectable candidates as calculated using the radio luminosity functions from \citet{2007MNRAS.375..931M}, which leads to  $\sim$3\% detection rate which \citet{2014MNRAS.440..696A} claim to be consistent with their number.
But \citet{2014MNRAS.440..696A} did not take into account the 10\% of redshift range found unusable by \citet{2011ApJ...742...60D}, which would move the detection rate closer to 4\%.
Therefore there is an indication that the detection rate of associated H{\sc i} absorption increases as the flux density limit for continuum sources is lowered.
Nevertheless, because of small number statistics we do not base our estimates on \citet{2011ApJ...742...60D} but on the detection rate from \citet{2014MNRAS.440..696A}.

BINGO will survey a 5000 deg$^2$ patch of the sky, lying in between HIPASS and the ALFALFA Pilot Survey coverages.
It will have much better sensitivity than either, while at the same time have a worse resolution (40\arcm~beam) than either.
The redshift range covered by it will be 0.12$<$z$<$0.48, higher than that covered by either survey.
This also implies that the potential number of AGNs will be larger than what will be calculated using the luminosity functions from \citet{2007MNRAS.375..931M}, as their fits were based on the Six-degree Field Galaxy Survey (6dFGS) matched with NVSS, and the galaxies have a median redshift of 0.043. 
We calculated the number of expected detections of H{\sc i} absorption for 5$\sigma$ detection of at least 30\% peak absorption against background continuum sources - in order to base our calculations on the numbers from \citet{2014MNRAS.440..696A}.
Therefore we are conservative regarding our estimates given the potential increase in detection rates with better continuum flux sensitivity discussed earlier.
As BINGO will cover a large redshift range and the sensitivity of the instrument will vary across the range, we do the calculations for four $\Delta$z~=~0.1 redshift slices.

Table~\ref{tab:bingo} presents the results of calculating the number of expected detections for BINGO for the different redshift slices.
Column 1 lists the mean redshift for each of these slices, and quantities like the luminosity distance for sources in each bin is calculated based on this mean redshift.
Although the actual thermal noise target for BINGO is lower, the relevant rms noise to consider is that due to the integrated HI emission from sources within the voxel being observed (the signal the experiment aims to detect), which is expected to be of the order of 0.1 mK, for which the corresponding 5$\sigma$ limits are given in column 2.
The continuum flux limits assuming that the absorption will be detected at 5$\sigma$ when having a minimum $\tau_{peak}$ of 0.3, are given in column 3 of Table 1.
We note that other continuum sources in the beam will not change these numbers - they will merely raise the level of the total continuum emission measured.
Columns 4 and 5 give the expected number of continuum sources above the flux limit for star forming galaxies and radio-loud AGNs respectively.
They are calculated using eqns. (5) and (6) from \citet{2007MNRAS.375..931M}.
As expected at these redshifts for the high power end of the radio luminosity function, AGNs dominate the number counts.
Finally, column 6 gives the expected number of detections in each redshift bin for a 2\% detection rate.

As we can see from Table 1, even with a conservative 2\% detection rate, and estimating source numbers based on a radio luminosity function which is biased towards low redshift sources and therefore does not take into account higher occurrences of AGNs with increasing redshift, we expect a large number of associated H{\sc I} absorption detections with BINGO.
A point to note here is given the typical velocity widths of 100 \kms\ or more for these absorption features, one should aim for channel widths no smaller than 10 \kms.
At the higher redshift end of the band this corresponds to around 30 KHz.
For detailed science regarding determining inflow / outflow velocities and source morphologies, each detection should ideally be followed up by high resolution interferometric observations.

According to the classification of H{\sc i} IM experiments by \citet{2015ApJ...803...21B}, the targeted IM experiment BINGO is a Stage I setup.
Such experiments are to be superseded by more sensitive experiments with wider sky coverage, through Stage II and III experiments.
With conservative assumptions, we expect BINGO to more than double the number of known detections of associated H{\sc I} absorptions.
Therefore detection and follow-up studies of associated H{\sc i} absorption will be an obvious avenue for commensal science using upcoming H{\sc i} IM experiments.

\end{appendix}

\label{lastpage}

\end{document}